\documentclass[fleqn,usenatbib, useAMS]{mnras}

\usepackage{newtxtext,newtxmath}
\usepackage[T1]{fontenc}
\usepackage{ae,aecompl}
\usepackage[utf8]{inputenc}
\usepackage{threeparttable}
\usepackage{booktabs, caption, makecell}
\usepackage{pdflscape}
\usepackage{geometry}
\usepackage{float}
\usepackage{pifont}
\usepackage{graphicx}	
\usepackage{amsmath}	

\newcommand{\Line}[3]{\Ion{#1}{#2}~#3\,\AA}
\newcommand{\Lines}[3]{\Ion{#1}{#2}\,#3\,\AA}
\newcommand{\Ion}[2]{#1{\,\sc#2}}

\usepackage{rotating}
\usepackage{float}
\usepackage{tabularx}
\usepackage{natbib}
\usepackage{hyperref}
\usepackage{academicons}
\usepackage{svg}
\newcommand{\orcid}[1]{\href{https://orcid.org/#1}{\includegraphics[width=10pt]{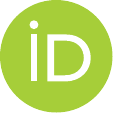}}}
\newcommand{\Teff}{\mbox{$T_{\mathrm{eff}}$}}
\newcommand{\logg}{\mbox{$\log g$}}
\newcommand{\Msun}{\mbox{$\mathrm{M_\odot}$}}

\newcommand{\gcolor}{\mbox{$G_{\mathrm{BP}}-G_{\mathrm{RP}}$}}
\newcommand{\av}{\mbox{$A_{V}$}}

\title[Ultraviolet extinction correlation with 3D dust maps]{Ultraviolet extinction correlation with 3D dust maps using white dwarfs}
 
\author[Sahu et al.]
{Snehalata Sahu \orcid{0000-0002-0801-8745}$^{1}$\thanks{E-mail: snehalatash30@gmail.com}, 
Pier-Emmanuel Tremblay$^{1}$, Rosine Lallement$^{2}$, Seth Redfield$^3$,  Boris T. G\"ansicke$^{1}$\\
$^{1}$ Department of Physics, University of Warwick, Coventry, CV4 7AL, UK\\
$^{2}$ GEPI, Observatoire de Paris, PSL University, CNRS, 5 Place Jules Janssen, 92190, Meudon, France\\
$^{3}$Astronomy Department and Van Vleck Observatory, Wesleyan University, Middletown, CT 06459-0123, USA
}

\date{Accepted 2024 October 07. Received in original form 2024 July 03}

\begin{document}
\label{firstpage}
\pagerange{\pageref{firstpage}--\pageref{lastpage}}
\maketitle

\begin{abstract}
Accurate astrometric and photometric measurements from \textit{Gaia} have led to the construction of 3D dust extinction maps which can now be used for estimating the integrated extinctions of Galactic sources located within 5\,kpc. These maps based on optical observations may not be reliable for use in the ultraviolet (UV) which is more sensitive to reddening. Past studies have focused on studying UV extinction using main-sequence stars but lack comparison with 3D dust maps. White dwarfs with well-modeled hydrogen-dominated (DA) atmospheres provide an advantage over main-sequence stars affected by magnetic activity. In this work, we study the variation of UV extinction with 3D dust maps utilising \textit{HST} and \textit{GALEX} observations of DA white dwarfs located within 300\,pc. We used \textit{HST} COS spectroscopic data of 76 sight lines to calculate the optical extinction from \ion{Si}{ii} column densities and validate our results with the kinematic model predictions of the local interstellar medium. Also, we combined \textit{GALEX} and \textit{Gaia} photometric observations of 1158 DA white dwarfs to study UV reddening by comparing observed and modeled colour-colour relations. We calculated \textit{GALEX} non-linearity corrections and derived reddening coefficients ($R(NUV-G)=6.52\pm1.53$ and $R(FUV-G)=6.04\pm2.41$) considering their variations with optical extinction ($\av<0.1$\,mag), and found them to be in good agreement with known extinction laws. \textit{HST} analysis suggests a positive bias of 0.01--0.02\,mag in the optical extinction from 3D maps depending on the Galactic latitude. These results independently confirm the validity of 3D dust maps to deredden the optical and UV observations of white dwarfs.
\end{abstract}


\section{Introduction}
Extinction provides a measure of the amount of light absorbed and scattered by dust grains as it traverses the interstellar medium (denoted by  $A_\mathrm{\lambda}$ in magnitude units for a specific wavelength). It is a wavelength-dependent phenomenon, with shorter wavelengths of light being scattered more than longer ones, thus causing the reddening of starlight (expressed as $E\mathrm{(\lambda-\lambda_{1}})=A_\mathrm{\lambda}-A_\mathrm{\lambda_{1}}$, where $\lambda_{1}$ is a reference wavelength). Generally, $E(B-V)$ based on the optical observations in standard Johnson $B$ and $V$ bands is used as a measure of reddening. The extinction laws \citep{seaton1979,Cardelli1989, Fitz1999} are characterised by the parameter $R(\lambda)$, termed as the extinction coefficient representing the ratio of total to selective extinction ($A_\mathrm{\lambda}/E(B-V)$). The reddening coefficient $R(\lambda-\lambda_{1})$ is also an indicator of the extinction law defined by the ratio $E\mathrm{(\lambda-\lambda_{1}})/E(B-V)$. Extinction and reddening corrections are essential for determining the intrinsic luminosities and colours of obscured astronomical objects such as stars and galaxies, thus carrying significant impact in almost all areas of modern astronomy. Traditionally, two-dimensional dust reddening maps \citep[e.g.,][]{schlegel1998, planck2014} have been widely used for these corrections, which are especially useful for extra-galactic sources. However, thanks to the accurate and precise parallax measurements from the \textit{Gaia} mission, complemented by ground-based photometric observations, advanced three-dimensional dust maps now trace the reddening both as a function of distance within the Milky Way and angular position in the sky \citep{green2019, rosine2019, leike2020, vergely2022, edenhofer2024}.

From extinction law studies \citep{Cardelli1989, Fitz1999, gordon2023}, it is well known that the interstellar extinction shows a steep rise in the ultraviolet (UV) characterised by a strong bump at around 2175\,\AA~\citep{Fitz1986}. Though the exact origin of this feature remains uncertain, the carbonaceous grains are often proposed to be the main carriers \citep{sorell1990, Li2001, bradley2005, steg2010}. Measurements of UV extinction are useful to investigate dust properties at different lines of sight, in particular very diffuse regions that are subjected to small optical extinction, such as high Galactic latitudes. Moreover, recent studies \citep[e.g.][]{butler2021} have shown that \textit{NUV} extinction is a more fundamental measure of dust column density than the optical reddening $E(B-V)$ or extinction (\av). Thus, studying UV extinction can provide better constraints on the size and density of the dust grains in different environments of our Galaxy.

UV extinction studies have gained more attention in the last two decades with the launch of the UV spacecraft, \textit{Galaxy Evolution Explorer} (\textit{GALEX}). Since its launch in 2003, \textit{GALEX} has conducted the largest ever UV imaging sky survey providing photometry of millions of hot sources in \textit{FUV}(1344--1786\,\AA) and \textit{NUV} (1771--2831\,\AA) bands with 4--5\,arcsec resolution \citep{martin2005}. With its primary focus on studying star formation and galaxy evolution, \textit{GALEX} photometric data has also been widely used in the characterisation of interstellar dust and extinction such as deriving UV extinction coefficients utilising vast samples ($\sim$million) of main sequence stars \citep{yuan2013, Sun2018, zhang2023} and generating UV extinction maps covering high Galactic latitude ($b\geq10^{\circ}$; \citealt{Sun2021}). However, these studies are mainly based on main-sequence stars with $\Teff\leq8000$\,K that possess magnetically active atmospheres resulting in hot spots or flares that can affect the ultraviolet photometry. Furthermore, \cite{yuan2013} reported large disagreement in \textit{FUV} extinction coefficient between the observations and theoretical predictions, suggesting the need for a hotter sample of main sequence stars or other spectral types.

Most white dwarfs possess a simple atmosphere that is dominated by hydrogen (spectral type DA at optical wavelengths). Their atmospheres are not magnetically active like those of main sequence stars, hence are advantageous to probe the UV extinction. Many of the white dwarfs with temperatures hotter than 10\,000\,K predominantly emit in the UV. \cite{Bianchi2011} has provided a catalogue of hot white dwarfs in the Milky Way demonstrating the sensitivity of \textit{GALEX} in detecting and characterising these hot stellar sources. More recent studies have used \textit{GALEX} white dwarfs to comment on the survey's flux calibration \citep{camarota2014, wall2019, bohlin2019} and the accuracy of white dwarf atmospheric parameters \citep{wall2023}. These studies also derive UV extinction coefficients $R(FUV)$ and $R(NUV)$ by comparing the observed \textit{GALEX} magnitudes to model atmosphere predictions \citep{wall2019}. However, past research did not compare the variation of \textit{GALEX} extinction coefficients with 3D optical dust maps.

White dwarfs have also significantly contributed to our understanding of the environment and structure of the local interstellar medium (LISM). The advantage of using white dwarfs for interstellar studies is that their spectra are generally featureless having a simple continuum. Compared to the optical, the UV spectrum contains stronger resonance absorption lines (\ion{P}{ii}, \ion{S}{ii}, \ion{Si}{ii}, \ion{Si}{iii}, \ion{C}{ii}, \ion{Fe}{ii}, etc.) that are more sensitive in probing the interstellar gas. These ISM lines are useful to detect the warm and highly ionised gases that help to understand the physical state of the LISM \citep{Lehner2003, seth2004, rosine2011} and calculate the amount of neutral hydrogen that directly correlates with dust \citep{Jenkins2009}. Studying the correlation between the presence of dust and gas, inferred from the absorption lines of UV spectra, provides an alternative way to photometry in constraining extinction. Specifically, the observation of individual sources lying at different lines of sight provides a direct way to gain insights into the variation of $R(\lambda)$ in different environments \citep{gordon2023}, thus giving a clearer idea about the structure and properties of dust grains. Additionally, interstellar lines in the UV serve as an important diagnostic tool to test the 3D dust-extinction maps that are based on optical photometric measurements \citep{green2019, rosine2022}. 

The ISM lines in the UV spectra of hot DA white dwarfs can be contaminated by photospheric lines, thus demanding high-resolution ($\Delta\lambda/\lambda\geq$ 90\,000) spectroscopic observations to study in detail the kinematics and structure of the LISM. For instance, observations from the Space Telescope Imaging Spectrograph (STIS) onboard \textit{Hubble Space Telescope (HST)} have been successfully used in resolving the narrow ISM lines and characterising the LISM \citep{seth2002, rosine2011}. With a primary motivation to identify the signatures of the accretion of planetary debris in hot white dwarfs, we have conducted a \textit{HST} snapshot survey acquiring intermediate-resolution ($\Delta\lambda/\lambda\approx$ 16\,000) spectroscopic data of 311 hydrogen-atmosphere DA white dwarfs (the largest sample so far) with the Cosmic Origins Spectrograph (COS; \citealt{sahu2023}). Taking advantage of the large sample size, we can now utilise these data to measure the interstellar gas column densities and investigate their connection with dust. 

This paper is divided into two objectives: 
first, we analysed ISM parameters towards 180 sight lines derived from \textit{HST} COS spectra of white dwarfs within 200\,pc, and infer their extinction from \ion{Si}{ii} column densities. We selected 76 sight lines with well-measured parameters and that satisfy a single component fit, to check the consistency with 3D extinction maps from EXPLORE \citep{vergely2022}. We also derived the extinction by measuring the dust depletion and hydrogen column densities of 17 targets using several UV lines in addition to \ion{Si}{ii}. We demonstrate the reliability of our results by comparing the observed ISM velocities with the predictions of a LISM kinematic model \citep{seth2008}. Secondly, we utilise 1158 spectroscopically confirmed DA white dwarfs lying within 300\,pc from the Sloan Digital Sky Survey (SDSS) with available \textit{GALEX} photometry, to derive the UV reddening coefficients in \textit{GALEX} \textit{FUV}and \textit{NUV} bands. We have investigated the correlation between UV extinction and 3D optical extinction, distance, and, Galactic latitude for sources with $T_{\rm eff}$ ranging between 13\,000--40\,000\,K. 

\section{\av~from 3D extinction maps}\label{sec:avmap}
For comparison with \av\ values determined from \textit{HST} and \textit{GALEX} data, we used the extinction values estimated by integrating within tridimensional (3D) extinction density maps, from the Sun to the white dwarf location throughout this work. 
We used the latest maps described in \cite{vergely2022}. They are based on the tomographic inversion of a large amount of extinction-distance pairs of individual targets, following a hierarchical technique described in \cite{rosine2019}. The latest maps used extinctions of $\simeq$ 35\,500\,000 individual stars, derived from \textit{Gaia} $G$, $G_{\rm BP}$, and $G_{\rm RP}$ and 2MASS $J$, $H$, and $K_{\rm S}$ photometric data, and their distances from \textit{Gaia} \,EDR3 parallaxes (dataset described in details in \citealt{rosine2022}). In addition to these data, accurate extinctions of an additional amount of $\simeq$ 6\,000\,000 stars derived from both spectroscopic and photometric measurements were also used. The two catalogs were inter-calibrated following a newly devised technique. The achievable spatial resolution of the maps is governed by the target star density and distribution, which is highly variable and depends primarily on the distance to the Sun. The hierarchical inversion technique ensures that the spatial correlation kernel is compatible everywhere with the local target density. For reasons linked with the computational time, several maps with different maximum spatial resolutions and distance coverages were built (see the EXPLORE website\footnote{\url{https://explore-platform.eu/}}). Here, we used a composite integration using all maps to benefit from the highest resolution everywhere along the path to the target. The integration starts within a 5\,pc resolution map and a maximum distance of 750\,pc from the Sun within the Galactic plane, and when (or if) the line-of-sight reaches the boundary of the map, we
continue integrating in the lower (10\,pc) resolution, more extended map, and so on. Uncertainties on the extinctions calculated with the 3D maps are spatially dependent. Uncertainties on the individual extinctions of the target stars used to build the maps vary from one source to the other, are on the order of $\simeq$ 0.02 mag in the best cases, and often reach 0.2 mag or more for distant targets. 

Biases are expected, even for the lowest extinctions, i.e. for stars at short distances and/or high latitudes. 
An attempt to estimate the bias at shorter distances/low
extinctions was made in \cite{vergely2022}, based on UV absorption data. They found an average positive bias of $\simeq$0.01\,mag on the map-integrated extinctions \av\ within the first 50\,pc. To correct for this bias, they subtracted this quantity from all individual measurements. However, because the 3D inversion uses extinction spatial gradients, and there is smoothing within the regularisation correlation kernel, such a correction may not be fully effective. Thus, the main goals of this work are to estimate the remaining bias and test the 3D dust maps using two independent methods: (1) ISM lines with \textit{HST} spectroscopy, and, (2) UV reddening utilising \textit{GALEX} photometric observations of white dwarfs.

\section{ISM analysis using HST spectroscopy}\label{sec:hst_ism}

\begin{figure*}
\centering
\includegraphics[width=\textwidth]{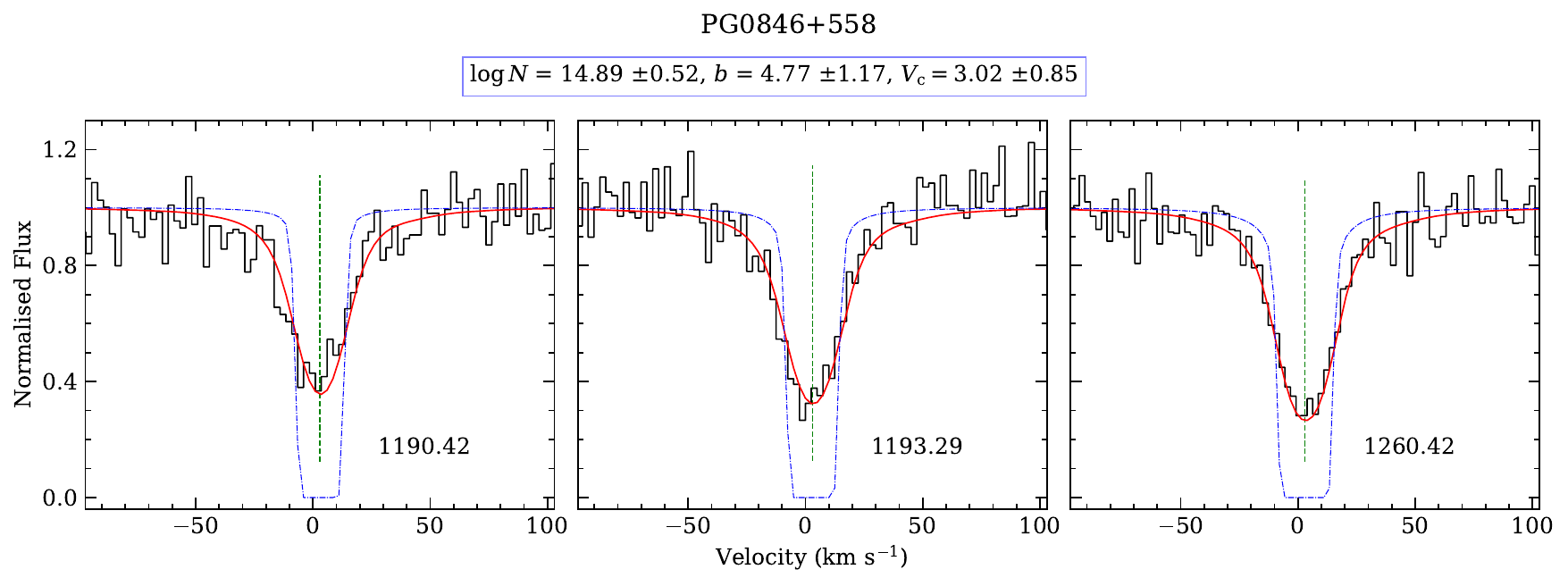}
\caption{\Ion{Si}{ii} absorption lines (normalised) observed by \textit{HST} COS for the white dwarf PG\,0846$+$558 are shown in black solid lines around the nominal vacuum wavelengths of 1190.42, 1193.29 and 1260.42\,\AA. The blue dash-dotted and red solid lines represent the best fit theoretical profiles to the observed spectra before and after convolution with COS line spread function. The green dashed line corresponds to the central velocity of the ISM cloud ($V_{\rm c}$ in km\,s$^{-1}$) obtained from the best-fit model. The model parameters are labeled at the top where log\,$N$ is the logarithm of column density and $b$ is the velocity dispersion in km\,s$^{-1}$.}

\label{fig:ism_fit_params}
\end{figure*}

\subsection{Sample selection}
The {\it HST} COS snapshot survey used in this study comprises 311 DA white dwarfs \citep{sahu2023}. In addition to this sample, we also included 12 DA stars that were observed after the published work in August 2023 during Cycles 29--30 (Program IDs 16642, 17420). The observations were obtained with COS grating G130M ($R\approx$ 16\,000) centered at wavelength 1291\,\AA. The wavelength coverage is 1130--1430\,\AA~with a gap at 1278--1288\,\AA\ due to the space between the two detector segments. We refer to \cite{sahu2023} for more details about the sample selection.

COS white dwarf spectra can have metal absorption lines (of varying strengths) arising in the photosphere that can mimic the ISM lines, therefore it is important to identify and disentangle them. The interstellar lines have velocities that can be similar or different to the photospheric lines, depending on the ISM cloud that the starlight has traversed along the line of sight. We identify several ISM lines in the COS spectra with the dominant ones being \ion{C}{ii} (1334.53, 1335.70\,\AA), \ion{O}{i} (1302.17\,\AA), \ion{N}{i} (triplet near 1200\,\AA), \ion{S}{ii} (1250.68, 1253.81, 1259.52\,\AA) and \ion{Si}{ii} (1190.42, 1193.29, 1260.42\,\AA). These species are well correlated with the gas column and are not significantly depleted onto dust grains \citep{Jenkins2009}. Among them, \ion{C}{ii} and \ion{S}{ii} transitions are the most affected by the photospheric lines, which are difficult to disentangle given the low resolution of COS. In addition, the \ion{C}{ii} lines are very broad and sometimes heavily saturated. In the case of other lines, \ion{O}{i} and \ion{N}{i} suffer from airglow contamination which can lead to inaccurate column density measurements. Hence, we focused on \ion{Si}{ii}\ lines for this study, where \Line{Si}{ii}{1260.42}\ is the strongest absorption line among other Si resonance lines seen in the spectra. To account for photospheric contamination, we visually inspected each spectrum for the presence of the \Ion{Si}{ii} doublet at 1265\,\AA\ that only has a photospheric origin (accretion of planetary material) and discarded them. Furthermore, we excluded the sources with detected circumstellar disks and known binaries such as double degenerates. Overall, we selected 173 DA white dwarfs with the detected \ion{Si}{ii}\ interstellar lines. We also included 7 magnetic white dwarfs (DAH spectral type) in the sample where the \Line{Si}{ii}{1260} line plausibly arises from the ISM leading to a total of 180 targets for further analysis.

\subsection{Methods}
The observed spectra were normalised by fitting a quadratic polynomial to the local stellar continuum around the three \ion{Si}{ii} lines (1190.42, 1193.29, 1260.42\,\AA). These absorption lines were masked to accurately determine the continuum. A width of 1\,\AA~corresponding to a velocity shift of $\approx$238\,km s$^{-1}$ was chosen for fitting. The reason behind the chosen width was to encompass both the absorption line and local continuum in the fitting. A Voigt profile was fitted to the normalised spectra following a standard technique as described in \cite{seth2004}. The free parameters of the model are: the central velocity of the interstellar cloud ($V_\mathrm{c}$ [km s$^{-1}$]), velocity dispersion or Doppler line width ($b$ [km s$^{-1}$]), and the ion column density ($N$ [in cm$^{-2}$]) which is a measure of the amount of material along the line of sight of the object. The rest wavelength, oscillator strength, and damping coefficient used in our fits are taken from \cite{morton2003}. The theoretical line profiles were convolved with the instrumental line spread functions (LSFs) following the procedure provided in \textit{HST} COS documentation\footnote{\url{https://github.com/spacetelescope/notebooks/blob/master/notebooks/COS/LSF/LSF.ipynb}}. 
A least-squares fit was performed using python package \textsc{scipy} based on trust region reflective algorithm. Since COS resolution is not high enough to resolve the individual cloud components of the LISM (with velocity differences $\leq$15\,km\,s$^{-1}$), as a simple approximation we assumed a single cloud component in the fitting.  The uncertainties in the best-fit parameters are determined from the covariance matrix. 

\subsubsection{Doppler width (b)}
An important parameter of the model, Doppler width ($b$) is related to the thermal temperature ($T$ [K]) and turbulent velocity ($\xi$ [km s$^{-1}$]) of the ISM by the following expression:
\begin{equation}
    b^2=\frac{2kT}{m}+\xi^{2}=0.016629\frac{T}{A}+\xi^{2}
    \label{eqn:doppler_width}
\end{equation}
where $k$ is Boltzmann's constant, $m$ is the mass of the observed ion and $A$ is the atomic weight of the element.
The LISM is warm and partially ionised with a temperature varying from 2\,000 to 12\,000\,K with an average of 6680 $\pm$ 1490\,K \citep{seth2004_2}. Adopting this temperature range and a mean value of 2.24 $\pm$ 1.03 km s$^{-1}$ for turbulent velocity $\xi$ in Eqn.\,\ref{eqn:doppler_width}, the $b$ value is expected to vary in the range 2.49--3.48 km s$^{-1}$ for Si. This criterion is valid only for cases where a single cloud is present in the line of sight. For the cases with multiple clouds, the $b$ values can be significantly larger ($>6$ km s$^{-1}$) than the expected values from LISM temperature. Given the COS medium-resolution spectra, there is a strong degeneracy between $N$ and $b$ for saturated lines when the theoretical spectrum goes to zero before the convolution with the COS LSF. Hence, it is challenging to constrain $b$ from any single transition of \ion{Si}{ii} that is expected to be in rough agreement with the LISM temperature. To overcome this and obtain accurate column densities, all three transitions of \ion{Si}{ii} with different oscillator strengths were fitted simultaneously. 

We fitted 180 stars in the sample inspecting each fit visually. The Voigt profile fit to the \ion{Si}{ii} absorption lines of one such white dwarf, for instance, is shown in Fig.\,\ref{fig:ism_fit_params}.

\subsection{Deriving \av~from column densities}\label{sec:av_nh_dep}
\cite{Jenkins2009} provided an indirect method to infer the neutral hydrogen column density denoted as $N$(H) when it is unknown or can not be measured directly from the observations. We followed their procedure to determine $N$(H) from $N$(\ion{Si}{ii}) as provided below:
\begin{equation}
\centering
\log N({\rm H})+F_{*}A_{\rm X}= \log N(\rm X)-\log(\rm X/{\rm H})_{\odot}-B_{\rm X}+A_{\rm X}z_{\rm X}
\label{eqn:nh_nsi}
\end{equation}
where, $A_{\rm X}$, $B_{\rm X}$, $z_{\rm X}$ are the element depletion parameters and $F_{*}$ is the line of sight depletion strength factor as defined in \cite{Jenkins2009}. Here, larger values of $F_{*}$ signify stronger depletion for all elements which in turn implies that a large amount of metals missing in the gas phase are incorporated in the dust grains.

For Si, we adopted the values $\log({\rm Si/H})_{\odot}+12=7.61$, $A_{\rm Si}=-1.136$, $B_{\rm Si}=-0.570$ and $z_{\rm Si}=0.305$ as reported in Table\,4 of \cite{Jenkins2009}.

\begin{figure}
\centering
\includegraphics[width=\columnwidth]{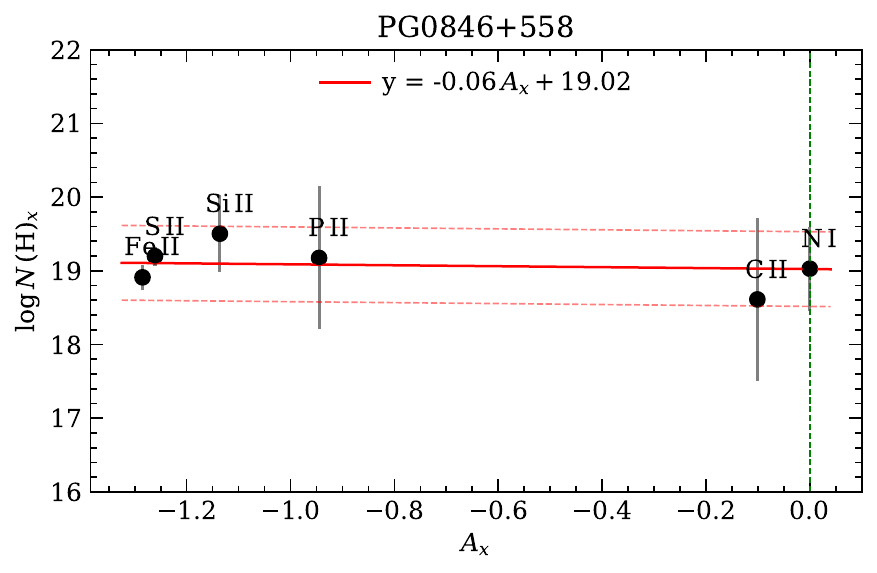}
\caption{Method for measuring synthetic depletion ($F^{*}$) and $N$(H) following \citet{Jenkins2009}. The $y$-axis corresponding to the right-hand side of Eqn.\,\ref{eqn:nh_nsi} for individual elements is plotted as a function of element depletion parameter $A_{\rm X}$. The red solid line shows the best-fit linear relation with the slope and intercept at $A_{\rm X}=0$ (green dotted line) corresponding to $F^{*}$ and $N$(H) respectively. The red dashed lines give the associated 1$\sigma$ uncertainties in the resultant column density measurements.}
\label{fig:dep_calc}
\end{figure}

To calculate synthetic $F_{*}$, a combination of elements that cover a wider range of $A_{X}$ values is required, which is not possible with the single element (Si). Hence, to address this, we checked for other lines such as \ion{C}{ii} (1334.53\,\AA), \ion{N}{i} (1199.55, 1200.22, 1200.71\,\AA), \ion{S}{ii} (1250.68, 1253.81, 1259.52\,\AA), \ion{P}{ii} (1152.17\,\AA), and \ion{Fe}{ii} (1144.48\,\AA). We excluded \ion{O}{i} (1302.17\,\AA) as the observations are affected by airglow contamination. As we mostly have a single transition for each element in the COS spectra, we place a limit on their $b$ values in the fit based on the $b$ obtained from the \ion{Si}{ii} lines (varying it within 1~sigma uncertainties) following Eqn.\,\ref{eqn:doppler_width}. We adopted the element depletion parameters provided by \cite{Jenkins2009} and followed the same fitting procedure as for Si with constraints on $b$ to determine the column densities. To obtain a large spread in $A_{\rm X}$ we require a combination of at least one lighter element (C, N) with heavy elements (Si, S, P, Fe). Hence, we chose only those targets satisfying these criteria. With the column densities of selected elements and following Eqn.\,\ref{eqn:nh_nsi}, a straight line was fitted using weighted least-squares method (giving weightage to the ions having lower uncertainties in column densities) to obtain synthetic $N({\rm H})$ and $F_{*}$. One example fit is shown in Fig.\,\ref{fig:dep_calc} for PG\,0846+558. 

The light elements (C, N) are important for measuring $F_{*}$ and $N$(H), however, these lines are heavily saturated in most of the cases leading to wrong column densities. Hence, it is difficult to determine $N({\rm H})$ from \ion{Si}{ii} column densities since the depletion measurements obtained from a straight line fit are unreliable. To address this, we first considered zero depletion ($F_{*}=0$) in Eqn.\,\ref{eqn:nh_nsi}, while calculating $N({\rm H})$ for the sources with well measured $N(\ion{Si}{ii})$. This is a reasonable approximation given an average depletion of $F_{*}\approx0.1$ noted for high Galactic latitude ($>50^\circ$) or nearby objects (within 100\,pc) based on 31 nearby white dwarfs located within the Local Bubble \citep{Jenkins2009}.

Finally, we derived $A_{V}^\ion{Si}{ii}$ from $N({\rm H})$ using the well-known gas-to-dust ratio relation \citep{Jenkins2009}:
\begin{equation}
\centering
A_{V}^\ion{Si}{ii}= 3.1 \times N({\rm H}) / 6\times10^{21}{\rm cm}^{-2}{\rm mag}^{-1}~
\label{eqn:av_nh}
\end{equation}
where the factor 3.1 is the average total-to-selective extinction ratio
$R(V)=A_{V}/E(B-V)=3.1$.

\subsection{Results}\label{sec:hst_res}
Among 180 stars, we selected 116 targets having a good fit with uncertainties less than 1 (log scale) in column density measurements of \ion{Si}{ii} and $b$ values $<$5 km s$^{-1}$ satisfying the single component fits. In the LISM, the multiple cloud components are common but not excessive. \cite{Malamut2014} has shown that a sight line within 100\,pc has, on average, 1--2 absorbers. Hence, our imposed criteria of $b$ values is reasonable. For the rest of the targets with larger $b$ values $>$5 km s$^{-1}$, we place a lower limit on the column densities. The lower column densities can arise plausibly from the line saturation due to stronger transitions, multiple cloud components being not resolved or hidden stellar contamination that cannot be evidenced from the weaker photospheric absorption line at 1265\,\AA. 
 
Based on the morphological studies of ISM clouds, \cite{seth2008} found that LIC and G clouds mostly dominate the sky in the LISM, while multiple clouds can be present in the boundaries especially in the regions with Galactic longitude between $40^{\circ}$ to $80^{\circ}$ and latitude between $-15^{\circ}$ to $+30^{\circ}$, and, between longitude $270^{\circ}$ to $320^{\circ}$ and latitude from $+20^{\circ}$ to $+50^{\circ}$ and $-70^{\circ}$ to $-30^{\circ}$. Hence, to investigate the targets having more than two absorbers along the line of sight, we compared their measured radial velocities with the predictions from \cite{seth2008}, who presented an empirical dynamical model of the LISM consisting of 15 warm clouds located within 15\,pc of the Sun. The model is based on the velocities of LISM absorbers obtained from high-resolution \textit{HST} spectra of 157 stars located up to 100\,pc from the Sun, i.e., the average distance to the boundary of the Local Bubble. This sample exhibits distinct kinematic clouds \citep{seth2015}.  The distribution of absorbers within the Local Bubble appears to be largely concentrated within about 15\,pc of the Sun based on the minimal rise in the number of detected absorbers as a function of distance \citep{seth2004,Malamut2014}. This distribution may be caused by the expanding shell of the latest supernovae explosions that have sculpted the Local Bubble over the last 10\,Myr \citep{zucker2024}. Extinction maps confirm the concentration of the LISM clouds within 15\,pc, a fairly sparse Local Bubble, and an abundance of interstellar clouds at the Local Bubble boundary beyond 100\,pc \citep{rosine2022}. Based on the \cite{seth2008} model, an online LISM kinematic calculator\footnote{\url{http://lism.wesleyan.edu/LISMdynamics.html}} is available that provides a list of clouds and their velocities that are predicted to traverse any given line of sight. We used this model to predict the radial velocities of each cloud along the line of sight for each object in our sample. Then, we compared the average radial velocities of all clouds for a given object ($V_{\rm LISM}$) with that measured from the \ion{Si}{ii} line ($V_\mathrm{c}^{\ion{Si}{ii}}$). 

Before comparing the observed and model velocities, we checked for discrepancies with COS wavelength calibration. The COS wavelength scale for FUV grating (G130M) is accurate to a velocity of 15\,km\,s$^{-1}$ which has been improved to 7.5\,km\,s$^{-1}$ with new disperison solutions for the observations later than 2018 \citep{plesha2018}. Hence, we imposed a criterion with the velocity difference to lie between $-12$ to $+12$\,km\,s$^{-1}$ that is roughly equal to the average of these values.
In addition, we chose only targets with $|A_{V}^{\rm map}-A_{V}^{\ion{Si} {ii}}| \leq 0.04$\,mag to avoid the sources having visibly poor fits ($\chi^{2}>3$). 
With all the criteria provided in Table\,\ref{tab:hst_sample}, we are left with a final sample of 76 targets for extinction comparisons. Out of this,  only 5 sight lines lie beyond 100\,pc, hence comparison with the \cite{seth2008} model is valid for our sample. The histograms depicting the \av\ and distance distribution of the selected sample are shown in Fig.\,\ref{fig:av_dist_hist}. Their best-fit parameters are provided in the appendix in Table\,\ref{tab:ism_param}. 

\begin{table}
\caption{Criteria for \textit{HST} sample selection for \av\ comparison study}
\centering
\begin{tabular}{cc}
\hline
criteria & Number\\\hline
COS sample & 323 DA, 7 DAH\\
ISM (no photospheric metals) & 180\\
error in log$N$(\ion{Si}{ii}) $\leq1$ & 156\\
$1\leq b$ [km/s] $\leq$5 and error in $b\leq3$ & 116\\
$-12\leq V_{\rm LISM}-V_{\rm c}^{\ion{Si} {ii}}$ [km/s] $\leq12$ & 86 \\
$|A_{V}^{\rm map}-A_{V}^{\ion{Si} {ii}}|$ [mag] $\leq0.04$ & \textbf{76} \\\hline
\hline
\end{tabular}
\label{tab:hst_sample}
\end{table}

\begin{figure}
\centering
\includegraphics[width=\columnwidth]{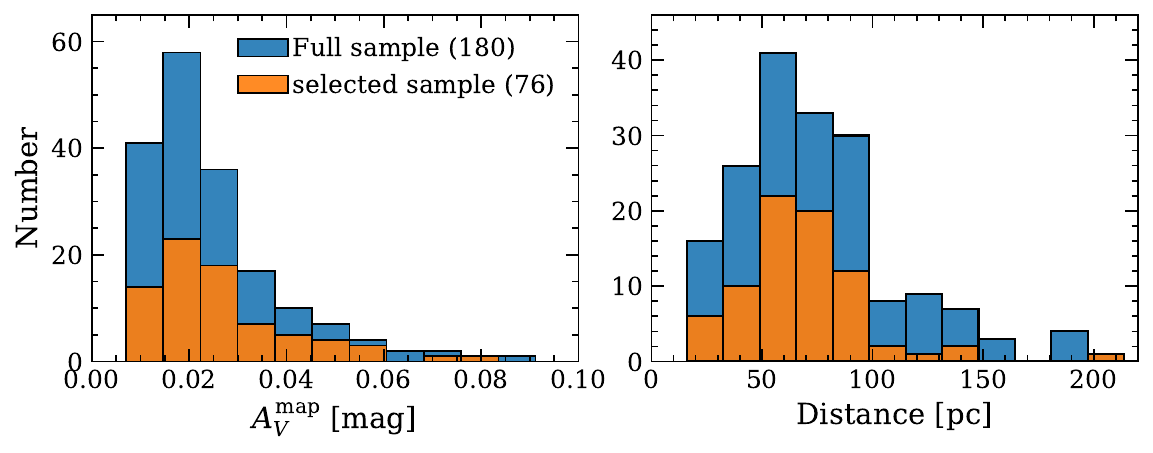}
\caption{Histograms showing the $A_{V}^{\rm map}$ and distance distributions of the full COS sample and selected 76 sight lines for our analysis.}
\label{fig:av_dist_hist}
\end{figure}

\begin{figure}
\centering
\includegraphics[width=\columnwidth]{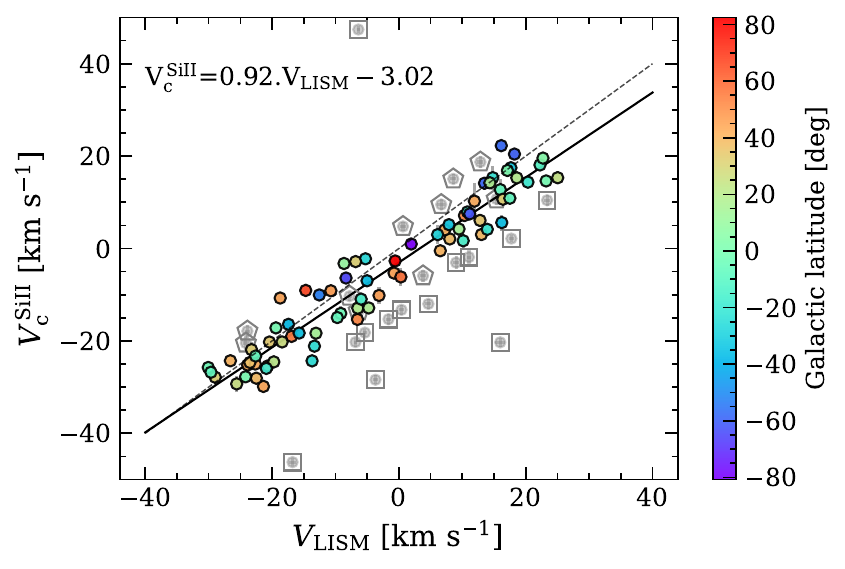}
\caption{Central velocities of the ISM measured from \Ion{Si}{ii} as a function of the velocities (mean) predicted from the LISM kinematic model \citep{seth2008}. The colourbar shows the Galactic latitude of the selected 76 targets for extinction comparisons (Table\,\ref{tab:hst_sample}). The solid black line denotes the best-fit linear relation, where 13 sources (grey squares) with $|V_{\rm LISM}-V_{\rm c}^{\ion{Si}{ii}}| > 12$\,km\,s$^{-1}$ and 10 sources (grey pentagons) with $|A_{V}^{\rm map}-A_{V}^{\ion{Si} {ii}}|>0.04$\,mag are excluded from the fit.}
\label{fig:vel_lism_comp}
\end{figure}

\begin{figure*}
\centering
\includegraphics[width=\textwidth]{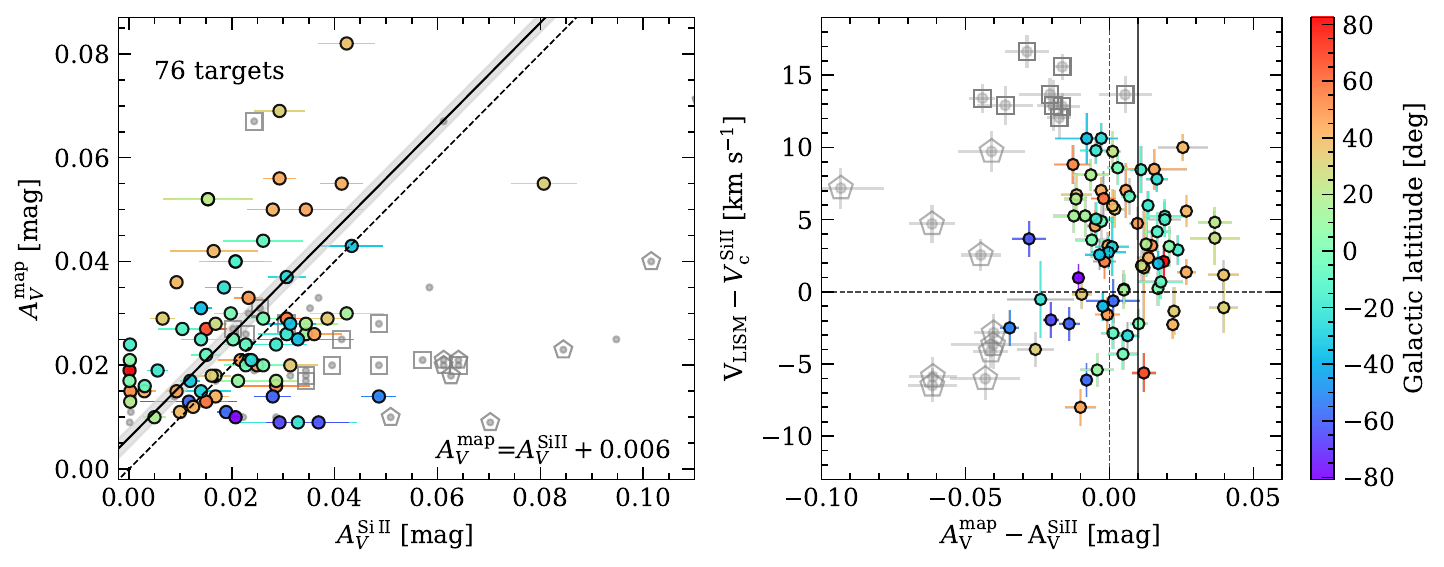}

\caption{{\it Left panel:} Comparison between extinctions measured from \ion{Si}{ii} (1190.42, 1193.29, 1260.42\,\AA) absorption lines ($A_{V}^{\ion{Si}{ii}}$) in the \textit{HST} COS spectra of 76 selected white dwarfs and obtained from the 3D extinction maps ($A_{V}^{\rm map}$). 17 targets marked in grey dots do not traverse through any clouds along the sight line. The black solid line in the left panel shows the best-fit intercept assuming a slope of 1, while the black dashed line denotes the case of one-to-one relation. {\it Right panel:} Their velocity difference as a function of the difference in \av~obtained from the 3D extinction map and those measured from \ion{Si}{ii}. 13 targets (grey squares) with $|V_{\rm LISM}-V_{\rm c}^{\ion{Si}{ii}}| > 12$\,km\,s$^{-1}$ and 10 sources (grey pentagons) with $|A_{V}^{\rm map}-A_{V}^{\ion{Si} {ii}}|>0.04$\,mag are excluded from the comparison analysis (refer Table\,\ref{tab:hst_sample} for selection criteria). }
\label{fig:ism_Av_comp}
\end{figure*}

To examine the targets selected for extinction comparisons in more detail, we plot the observed line velocities as a function of the model velocities in Fig.\,\ref{fig:vel_lism_comp}. A straight line fit to the velocities resulted in:
\begin{equation}
   V_{\rm c}^{\ion{Si}{ii}}=0.92\,(\pm0.01)\times V_{\rm LISM}-3.02\,(\pm0.11) 
\label{eqn:v_comp}
\end{equation}
suggesting that the velocities are in general agreement given the $1\sigma$ uncertainties. 
\cite{seth2008} suggested a minimum uncertainty of 3\,km s$^{-1}$ in the observed velocity from medium-resolution data which could be the reason behind the observed bias. We note that 13 excluded sources (grey squares in Fig.\,\ref{fig:vel_lism_comp}) having larger velocity differences ($>12$ or $<-12$ km s$^{-1}$) lie far from the straight line fit of the velocity difference. 
Visually inspecting the fits of 10 other excluded sources that satisfy velocity criteria but have large \av\ differences ($|A_{V}^{\rm map}-A_{V}^{\ion{Si}{ii}}|>0.04$\,mag), we note that Voigt profile fits to the observed spectra are poor ($\chi^2>3$) due to asymmetric line profiles or marginally resolved multiple components.
These deviations indicate that a single component fit might not be valid in such cases due to the presence of several clouds at different velocities affecting the column density measurements. To address this problem, we need a multiple component fit which is only possible with high-resolution ($\Delta\lambda/\lambda\geq1\,00\,000$) spectroscopic observations at UV wavelengths \citep{seth2004}.

We selected 76 sources for extinction comparisons. The plot of $A_{V}^{\rm map}$ as a function of $A_{V}^\ion{Si}{ii}$ derived from $N(\ion{Si}{ii})$ and assuming zero depletion is shown in Fig.\,\ref{fig:ism_Av_comp}. Assuming a slope of one, we performed a straight line fit to the selected targets weighting in terms of the $\chi^{2}$ obtained from the ISM model fits. The resulting fit suggests that the extinction measurements from 3D maps have a positive bias of $\approx$0.01\,mag ($\pm0.002$) when compared to \av~from \ion{Si} {ii}. We find that the $A_{V}^\ion{Si}{ii}$ values for targets located at low Galactic latitudes ($-20^{\circ}$ to $+20^\circ$) are evenly distributed around the noted bias, while the sources at high Galactic latitudes ($\geq+30^\circ$) suggest a higher $A_{V}^{\rm map}$ bias of 0.02--0.03\,mag (right panel of Fig.\,\ref{fig:ism_Av_comp}). These sources have \ion{Si}{ii} velocities ranging from $\approx-22$ to $-30$\,km\,s$^{-1}$ and are located at Galactic longitudes between $1-50^{\circ}$. 

\begin{figure}
\centering
\includegraphics[width=0.9\columnwidth]{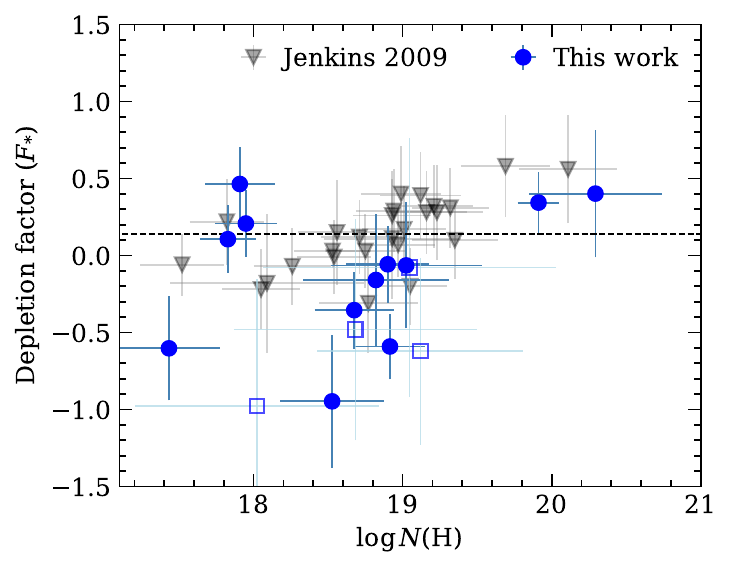}
\includegraphics[width=\columnwidth]{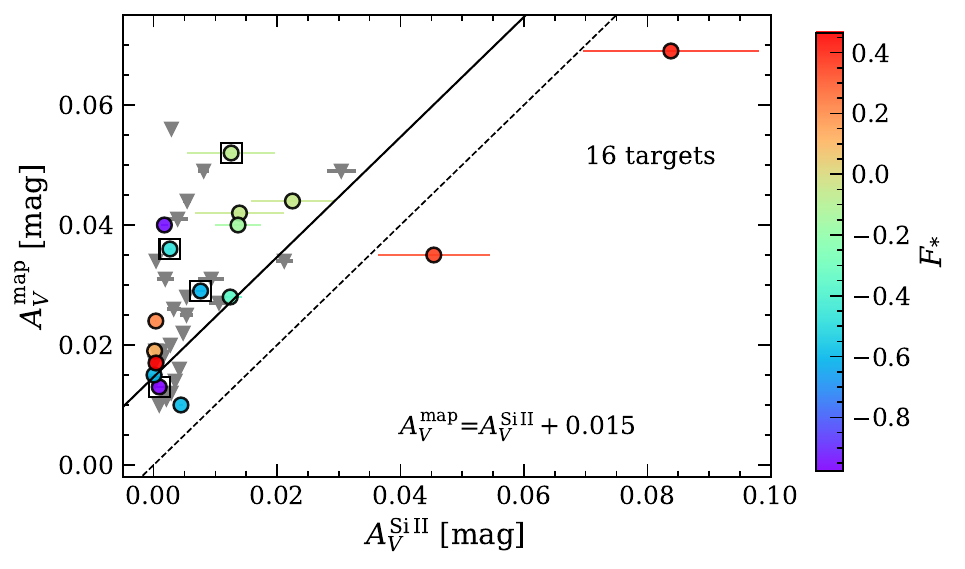}
\caption{\textit{Top panel:} Depletion ($F_{*}$) as a function of hydrogen column densities computed in this work for 12 targets (blue dots). \citet{Jenkins2009} measurements for 31 nearby white dwarfs ($<200\,$pc) are marked in grey lower triangles for comparison. 
\textit{Bottom panel:} Taking the measured depletion into account, the $A_{V}^{\ion{Si}{ii}}$ values are derived where the colour bar represents $F_{*}$. Similarly, the $A_{V}^{\ion{Si}{ii}}$ values are calculated from $N(\ion{Si}{ii})$ measurements using the depletion measurements of 31 white dwarfs from \citet{Jenkins2009}. The four targets having large uncertainties in $F_{*}>0.6$ are marked in squares in both plots.}
\label{fig:ism_Av_comp_dep}
\end{figure}

To investigate the effect of depletion, we derived the synthetic $F_{*}$ and log\,$N$(H) considering other UV absorption lines (C, N, S, P, Fe) following the procedure mentioned in Sec.\,\ref{sec:av_nh_dep}. Similar to \ion{Si}{ii} lines, we chose only those elements that have uncertainties $\leq1$ in column density measurements in log scale. In addition, we checked if their velocities are consistent with those derived from \ion{Si}{ii} lines ($<12$ or $>-12$\,km\,s$^{-1}$). Taking this into account, we found only 12 targets with reliable measurements (having uncertainties $\leq0.5$ in $F_{*}$ and $\leq1$ in log\,$N$(H)). The observed velocities of \ion{C}{ii} and \ion{N}{i} are found to be in excellent agreement with $V_\mathrm{c}^\ion{Si}{ii}$ with the differences ranging from only $-3$ to $+3$\,km\,s$^{-1}$. We have four targets with higher uncertainties in $F_{*}$ ($>0.6$ and $\leq1$) which are arising due to poor S/N and hence inaccurate column densities. Figure\,\ref{fig:ism_Av_comp_dep} (top panel) shows the distribution of $F_{*}$ as a function of log\,$N$(H) for these 16 targets along with \cite{Jenkins2009}. These targets are located within the Local Bubble (except one at a distance of $\approx$200\,pc). We find that their synthetic log\,$N(\rm H)$ range from 17.43 to 20.3 and $F_{*}$ from $-0.97$ to 0.46 being in general agreement with the values noted by \cite{Jenkins2009} given the one sigma uncertainties. For instance, \cite{Jenkins2009} reported lighter depletion in the direction of the south Galactic pole and higher depletion for stars lying between longitudes $60^{\circ}$ to $100^{\circ}$ and at varying latitudes, which is also noticed in our targets lying at similar distances. However, UV spectroscopic observations with higher resolution and S/N are required to improve these measurements and confirm the noticed trend.

Taking into account the depletion measurements in the calculations of log\,$N$(H) from log\,$N$(\ion{Si}{ii}), we compared the $A_{V}^{\rm map}$ as a function of $A_{V}^\ion{Si}{ii}$ for 16 sight lines as shown in  Fig.\,\ref{fig:ism_Av_comp_dep} (bottom panel). We find a positive bias of 0.015\,mag in $A_{V}^{\rm map}$ consistent with the case of 76 targets when considering an average depletion of $\approx0.14$  \citep{Jenkins2009}. We have 8 targets in total from \cite{Jenkins2009} and this work that disagree with $A_{V}^{\rm map}$ (difference larger than 0.03\,mag). We have compared them with a higher resolution 3D dust extinction map from \cite{leike2020} to check if this is a real discrepancy which is described later in the discussion.

The $A_{V}^{\ion{Si}{ii}}$ values were derived assuming an $R(V)$ value of 3.1 (see Eqn.\,\ref{eqn:av_nh}) and the average dust to gas ratio, represented by the reddening to H column ratio, and for which $E(B-V)= 1$ corresponds to a H column of $6\times10^{21}{\rm cm}^{-2}$. However, these values can vary significantly depending on the line of sight (los). If the variability of our results is mainly due to los to los variations of these quantities, and assuming zero depletion, one can examine their amplitudes:
if $R_{V}^{\rm los}$ is the total-to-selective extinction ratio for a given los, and $D/G^{\rm los}= E(B-V)\times 6\times10^{21}/ N({\rm H}) $ the normalized dust to gas ratio, then one can define: 
.
\begin{equation}
\centering
F^{\rm los}= A_{V}^{\rm map} / A_{V}^{\rm SiII} \times 3.1 = R_{V}^{\rm los} \times D/G^{\rm los} 
\label{eqn:av_nh_app}
\end{equation}

\begin{figure}
\centering
\includegraphics[width=0.9\columnwidth]{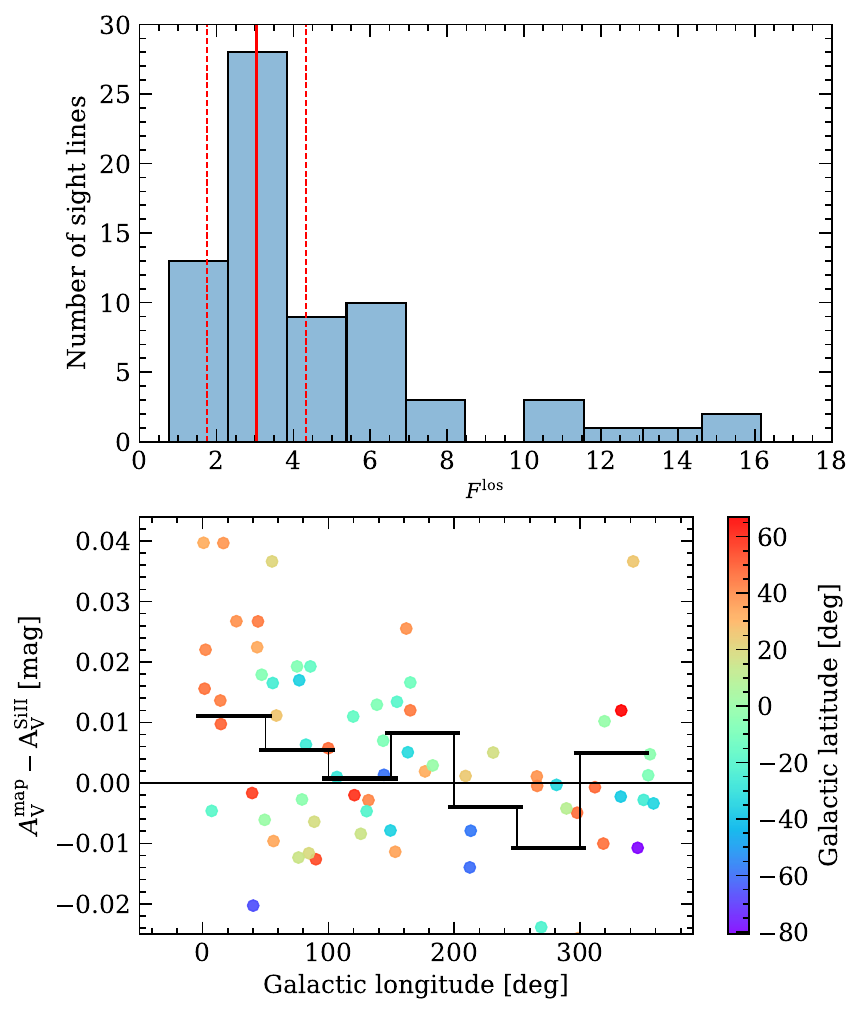}
\caption{Histogram of $F^{\rm los}$  for 76 sight lines in the top panel. The median of $F^{\rm los}$  for sources ranging between 2 to 6 is shown in a solid red line with the 1$\sigma$ errors in dashed red lines. The variation of $A_{V}^{\rm map}-A_{V}^{\ion{Si}{ii}}$ with the Galactic longitude is shown in the bottom panel while the Galactic latitude is shown in the colour bar. Higher \av\ difference implies a sight line with larger $F^{\rm los}$  value. The average of \av\ differences in each longitude bin (bin width of 50$^{\circ}$) is shown in black horizontal lines.} 
\label{fig:rv_galc}
\end{figure}

The histogram of \textbf{$F^{\rm los}$} values is shown in Fig.\,\ref{fig:rv_galc} (top panel). 
The column densities of $N(\ion{Si}{ii})$ for 76 sources range from $0.01-3.8\times10^{15}$\,ions \,cm$^{-2}$ corresponding to $N(\ion{H})$ of $0.02-15.6\times10^{19}$\,ions\,cm$^{-2}$. We find that 62~per cent of the targets have $F^{\rm los}$ values ranging from 2 to 6 with a median of 3.3 ($\pm1.2$) which is in good agreement with the mean value of 3.1 for the Milky Way based on the previous measurements and a \textit{classical} dust to gas ratio. For the targets with high $F^{\rm los}$  values ($>$6), the hydrogen column densities (based on \ion{Si}{ii}) are lower compared to those expected from the $A_{V}^{\rm map}$. There could be several reasons behind the large $F^{\rm los}$  values such as poor signal-noise ratio, inaccurate \av\ from the 3D maps, the gas-to-dust ratio being different from its mean value, or the assumption of zero depletion might not be valid for these sight lines which needs further examination with high-resolution spectroscopic data.

To study potential wide-area variations of differences between dust and gas, we plotted the \av\ difference as a function of Galactic longitude in Fig.\,\ref{fig:rv_galc} (bottom panel) where the targets with larger \av\ difference correspond to high $F^{\rm los}$  values. Considering a bin width of 50$^{\circ}$ in Galactic longitude, we calculated the average of $A_{V}^{\rm map}-A_{V}^{\ion{Si}{ii}}$ difference in each bin. We notice that the \av\ difference is larger ($\approx$0.01\,mag) in lower longitude bins and becomes negative (average is $\approx-$0.004 to $-0.01$\,mag) as we reach towards the targets located between longitudes 200--300$^{\circ}$. These regions corresponding to lower $F^{\rm los}$ values suggest a higher gas-to-dust ratio (see Sec.\,\ref{sec:discuss}).

\section{UV Reddening with \textit{GALEX}$-$Gaia photometry}\label{sec:galex}
To estimate reddening, and study its variation with distance and optical extinction, we combined \textit{GALEX} photometry with \textit{Gaia}. We focused on H-atmosphere DA white dwarfs for our analysis as they are the most numerous and their atmospheric opacities from UV to optical are well understood.

Before utilising the \textit{GALEX} sample for reddening study, we need to calibrate the observations to account for any deviations from other data sets or the white dwarf model colours. The magnitude and colour calibration along with the method and sample selection for reddening study are described in the following sections.

\subsection{GALEX calibration}\label{sec:galex_calib}
\textit{GALEX} magnitudes typically brighter than 17\,mag suffer from non-linearity issues and therefore have to be corrected before calibrating the colours. Past studies \citep{camarota2014, wall2019} have provided the linearity corrections based on white dwarf optical spectroscopic parameters (valid for magnitudes $\leq17\,$mag) which can be less reliable in the UV due to systematic uncertainties in the derived parameters \citep{sahu2023}. Hence, for this work, we recomputed the corrections based on UV white dwarf spectroscopic parameters of the COS sample \citep{sahu2023}. This is more appropriate for \textit{GALEX} bands which aim to place its flux scale on \textit{HST} photometric standards \citep{bohlin2019}. 

\begin{figure}
\centering
\includegraphics[width=\columnwidth]{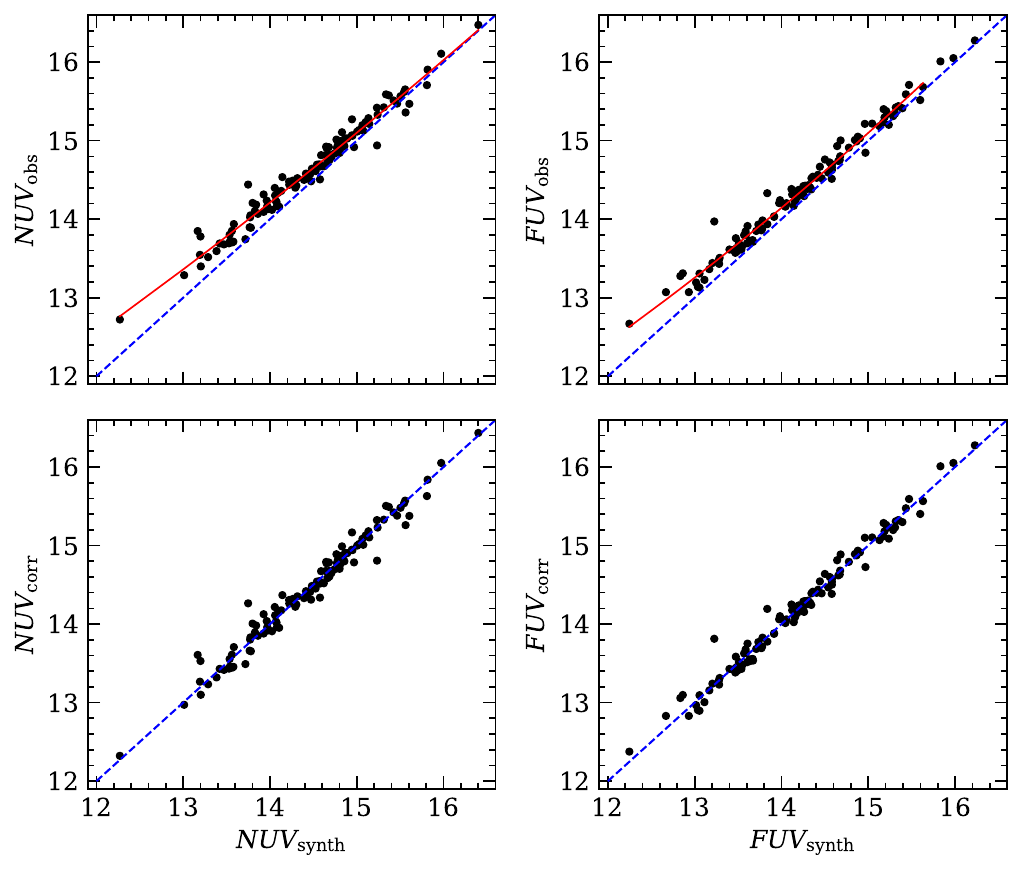}
\caption{Non-linearity correction to \textit{NUV} (left panels) and \textit{FUV} (right panels) magnitudes of \textit{GALEX} respectively, based on 114 DA white dwarfs from the COS sample lying within 100\,pc. The red solid lines in the top panels show the best-fit polynomial while the blue dashed lines in all panels show the expected one-to-one correlation. The corrections were applied in the observed magnitudes to bring them in agreement with the synthetically computed magnitudes (bottom panels).}
\label{fig:galex_nonlin_correc}
\end{figure}

Since extinction is less severe in the 100\,pc region within the Local Bubble \citep{rosine2003} in the interstellar medium, we utilised 114 DA white dwarfs located in this region that have both COS spectrophotometry \citep[excluding binary candidates;][]{sahu2023} and \textit{GALEX} photometry. The models corresponding to the COS white dwarf parameters \citep{sahu2023} and including extinction \citep{nicola2021} were convolved with the \textit{GALEX} filters to calculate synthetic magnitudes ($M_{\rm synth}$) in each band using python package \textsc{pyphot} \citep{pyphot}. These magnitudes were then compared with the observed magnitudes ($M_{\rm obs}$) and a quadratic polynomial was fitted to derive the coefficients ($c_{0}$, $c_{1}$, and, $c_{2}$) in each filter as shown in Fig.\,\ref{fig:galex_nonlin_correc} using the following expression:
\begin{align}
    M_{\rm obs} = c_{0} + c_{1}\times M_{\rm synth} +c_{2}\times M_{\rm synth}^{2}
\end{align}
The above-fit results were used to calculate the inverse quadratic coefficients ($q_{0}$, $q_{1}$, and $q_{2}$) as per the following expression:
\begin{align}
    M_{\rm corr} = q_{0} + (q_{1}\times M_{\rm obs} +q_{2})^{1/2}
\end{align}
where $M_{\rm corr}$ refers to the corrected magnitudes for non-linearity. After correction, the mean of the distribution of observed and synthetic magnitude differences in both the \textit{GALEX} bands are found to lie within their 1$\sigma$ photometric uncertainties ($\approx$0.01\,mag). Our corrections provided in Table \,\ref{tab:galex_mag_cor} are valid for magnitudes fainter than 13\,mag in both bands and brighter than 16\,mag in the \textit{FUV} and 16.5\,mag in the \textit{NUV} band.

\subsection{Method for reddening study}\label{sec:method}
Past \textit{GALEX} white dwarf studies have compared UV and optical photometry using individual white dwarf parameters, found either through photometric or optical spectroscopic fitting \citep{wall2019,wall2023}. Here we use a different approach by comparing the observed and predicted white dwarf colour loci, similar to what was done in \citet{Lopez2019} for J-PLUS photometric calibration. This procedure does not assign \Teff\ and \logg\ to individual white dwarfs, but instead uses model spectra to fit the observed colour-colour diagrams, allowing for possible photometric calibration offsets as free parameters. The advantage of this technique is that continuum hydrogen opacities are very well constrained for DA white dwarfs with 13\,000--40\,000\,K \citep{Saumon2022}, hence predicted colour-colour diagrams are expected to be robust. 

To measure the amount of reddening in the UV bands of \textit{GALEX}, we additionally rely on \textit{Gaia} $G$ magnitude and \gcolor~colour, which we assume have no zero point offsets as they have been shown to be in good agreement with other optical photometry \citep[see, e.g.,][]{McCleery2020}. The colours $FUV-G$ and $NUV-G$ giving a large baseline were selected and compared with \textit{Gaia} \gcolor\ having a good precision. We used the NLTE model spectra grid of \cite{Tremblay2011} with pure hydrogen composition applicable to DA white dwarfs. The grid is valid in the range $1500\leq\Teff\leq140\,000$\,K and $6.5\leq\logg\leq9.5$. A model of mass $\approx$0.6\,\Msun, which is representative of the majority of the white dwarf population, was used for comparison with the observed colour distribution. 

\begin{table}
\caption{Fitting parameters for non linearity correction in \textit{GALEX} bands valid for magnitudes $\geq$13\,mag and $\leq$16\,mag.}
\centering
\begin{tabular}{ccc}
\hline

coefficients & \textit{FUV} & \textit{NUV}\\\hline
\multicolumn{3}{c}{polynomial fit parameters}\\\hline
$c_{0}$ & 7.61 & 5.73\\
$c_{1}$ & 0.012 & 0.342\\
$c_{2}$ & 0.032 & 0.019\\
range & $\leq16$ mag & $\leq 16.5$  mag\\\hline
\multicolumn{3}{c}{inverse quadratic corrections}\\\hline
$q_{0}$ & $-$7.409 &  $-$21.38\\
$q_{1}$ & 45.256  &  78.3927\\
$q_{2}$ & $-$181.906 & 138.853\\\hline
\hline
\end{tabular}
\label{tab:galex_mag_cor}
\end{table}

\subsubsection{Colour calibration}\label{subsec:col_calib}
\begin{figure}
\centering
\includegraphics[width=\columnwidth]{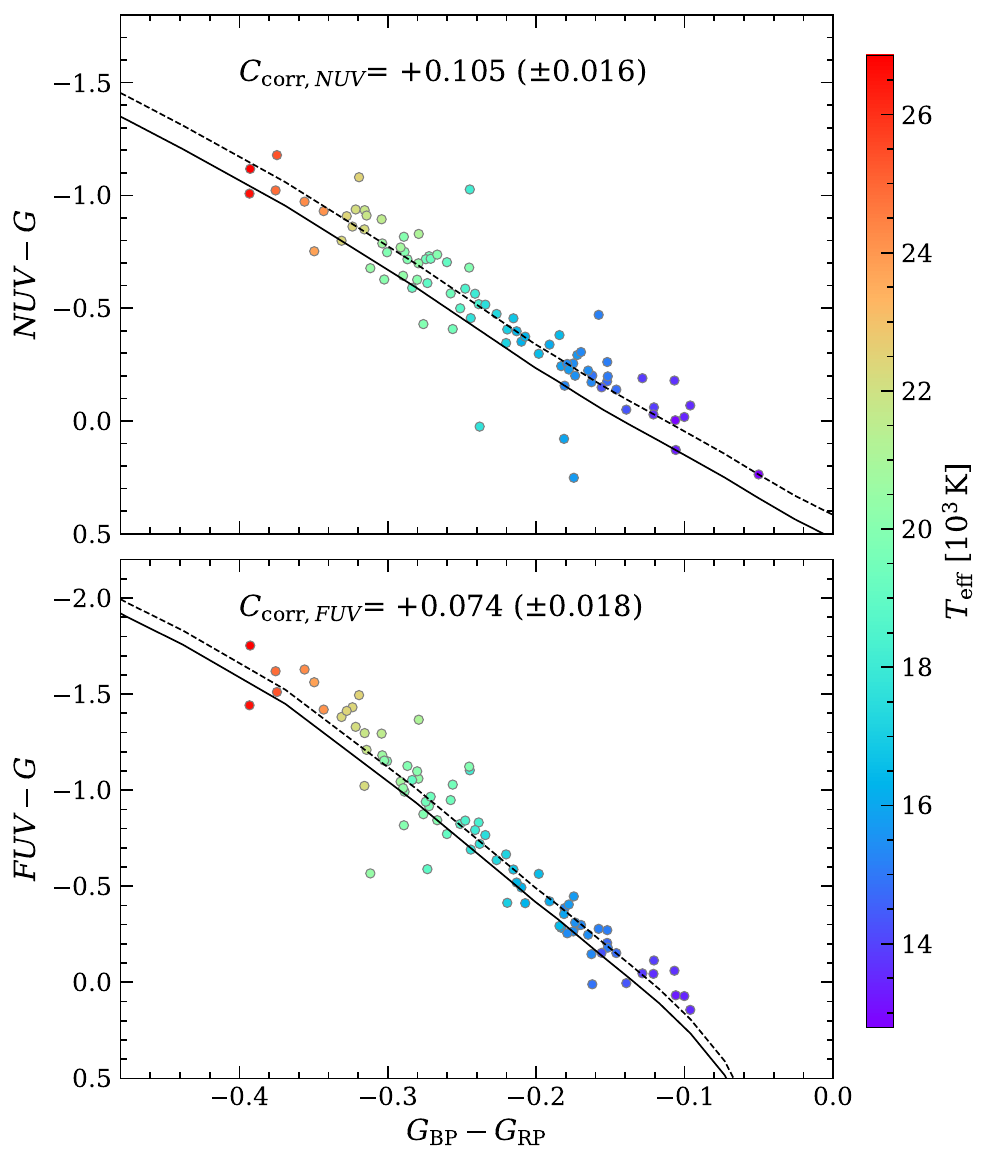}
\caption{\textit{GALEX-Gaia} colours as a function of only \textit{Gaia} colour (\gcolor) for 86 DA white dwarfs from \textit{HST} COS sample lying within 80\,pc. The \textit{GALEX} magnitudes are corrected for non-linearity following the method explained in Sec.\,\ref{sec:galex_calib}. Note that the plotted observed colours are dereddened using \citet{gordon2023} extinction law and $A_{V}^{\rm map}$ values. The black solid line represents predicted colours from the model spectra of \citet{Tremblay2011} for $0.6\,\Msun$ while the black dashed line corresponds to the shift in the median of the observed white dwarf distribution. These shifts define the $C_{{\rm corr}, \textit{NUV}}$ and $C_{{\rm corr}, \textit{FUV}}$ colour corrections (see Eqns.\,\ref{eqn:redd_calc1}-\ref{eqn:redd_calc2}) that we apply to the observations to match the white dwarf models, and that are identified on the panels. The colourbar represents the \Teff\ values obtained from \textit{Gaia} photometric fits with pure H models \citep{nicola2021}.}
\label{fig:mod_correc}
\end{figure}

Applying the non-linearity corrections, the corrected \textit{GALEX} magnitudes were used for calibrating the colours to model spectra.
We used 86 targets from the COS survey located within 80\,pc to define the colour corrections. We restricted the sample to close distances to minimise reddening effects. The corrections in the colours $FUV-G$ and $NUV-G$ were calculated by using a $\chi^{2}$ minimisation technique that best matches with the median of the observed white dwarf distribution as shown in Fig.\,\ref{fig:mod_correc}. To overcome the reddening effects in the calibration, the $G_{\rm BP}-G_{\rm RP}$ and $NUV/FUV-G$ colours were dereddened using $A_{V}^{\rm map}$ values and coefficients based on extinction law from \cite{gordon2023}. Then, the data were binned in $G_{\rm BP}-G_{\rm RP}$ colour ($x$-axis of Fig.\,\ref{fig:mod_correc}) and the median in each bin (to avoid the outliers) was calculated to find the locus. A linear interpolation was performed to compute the model colours in $FUV-G$ and $NUV-G$ corresponding to the observed and binned $G_{\rm BP}-G_{\rm RP}$ colour, which are then compared with the observed \textit{GALEX--Gaia} colours to determine the shift from a best-fit. The corrections to observed \textit{GALEX} magnitudes were found to be $+0.105$\,mag ($C_{{\rm corr}, NUV}$) and $+0.074$\,mag ($C_{{\rm corr}, FUV}$) in \textit{NUV} and \textit{FUV} bands respectively.

\begin{figure}
\centering
\includegraphics[width=\columnwidth]{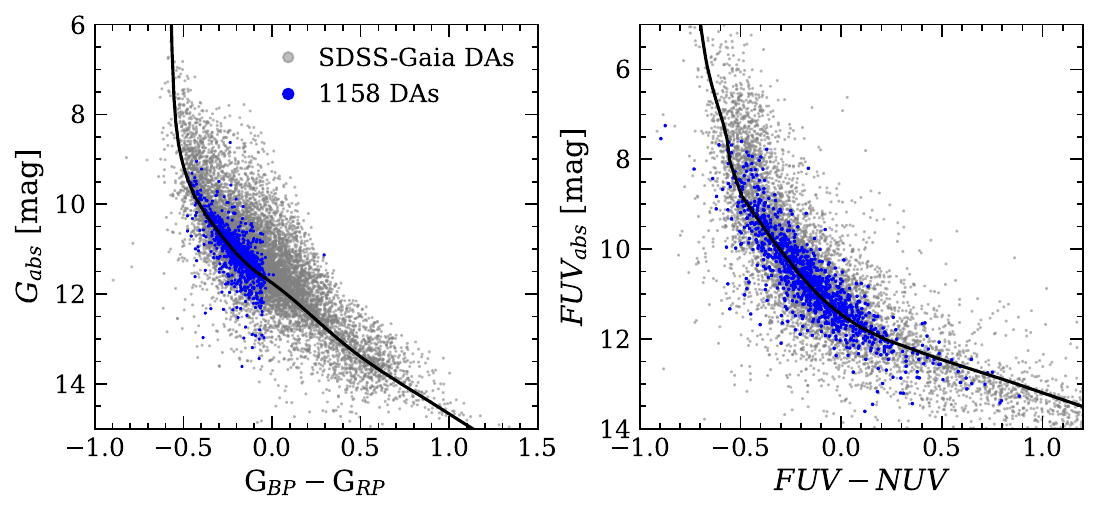}
\caption{\textit{Gaia} (left) and \textit{GALEX} (right) colour-magnitude diagrams of 11\,775 DA white dwarfs with spectral types based on SDSS spectra \citep{nicola2021} are shown in grey where the \textit{FUV} and \textit{NUV} magnitudes are corrected for non-linearity. The final sample of 1158 DA white dwarfs selected for reddening analysis is shown in blue. The white dwarf model fluxes \citep{Tremblay2011,Bedard2020} with a pure hydrogen atmosphere and a mass of 0.6\,\Msun\ are overplotted in a black solid line for reference.}
\label{fig:cmd_sample}
\end{figure}

\subsection{Sample selection}\label{sec:galex_sample}
We selected DA white dwarfs with $P_{\rm WD} \geq$ 0.75 from the SDSS-\textit{Gaia}\,EDR3/DR3 white dwarf catalogue of \cite{nicola2021} and cross-matched those with the \textit{GALEX} All Sky (AIS) survey (search radius of 3\,arcsec) resulting in 11\,455 objects. Among this sample, we chose sources brighter than 19\,mag in all \textit{Gaia} bands to lessen the scatter in \gcolor~as the photometry is unreliable close to the \textit{Gaia} magnitude detection limit (G $\approx$ 20.7\,mag) due to large ($\approx$10\,per cent) photometric flux uncertainties. Furthermore, we chose only sources lying in the \Teff~range 13\,000--40\,000\,K determined from the model colours based on \textit{Gaia} photometry \citep{nicola2021}. Note that this is the only step in this work where we have used individual white dwarf \Teff\ or \logg\ values. Above 40\,000\,K, both UV and optical colours become increasingly insensitive to $T_{\rm eff}$, hence also to photometric reddening. Below $\approx$ 13\,000\,K, white dwarf synthetic UV fluxes involve more uncertain Lyman $\alpha$ collision-driven quasi-molecular and extended red-wing opacities \citep{Allard1994,Kowalski2006,Saumon2014,OBrien2024}. Cool white dwarfs also have smaller UV fluxes, hence larger uncertainties associated with the \textit{GALEX} photometry.
We therefore added a cut-off of 20.5\,mag in \textit{FUV} and \textit{NUV} bands which is close to the limiting magnitude of the \textit{GALEX} photometric survey. Applying all the selection criteria (Table\,\ref{tab:galex_sample}), we are left with 1158 sources with $A_{V}^{\rm map}\leq0.1$\,mag and lying within 300\,pc for our reddening study in the UV. Figure\,\ref{fig:cmd_sample} shows the selected sample overplotted in \textit{Gaia} and \textit{GALEX} colour-magnitude diagrams in absolute scale.

Finally, the data were binned in distance or optical extinction, and the reddening in each bin was estimated using the following expressions:
\begin{align}
    E(NUV-G)=(NUV-G)_{\rm obs} +C_{{\rm corr}, NUV} -(NUV-G)_{\rm mod}\label{eqn:redd_calc1}\\
    E(FUV-G)=(FUV-G)_{\rm obs}+C_{{\rm corr}, FUV} -(FUV-G)_{\rm mod}\label{eqn:redd_calc2}
\end{align}
where "obs" and "mod" in subscript denotes the observed (corrected for non-linearity) and model colours, respectively, for the corresponding $G_{\rm BP}-G_{\rm RP}$ colours dereddened using $A_{V}^{\rm map}$ as in  Sec.\,\ref{subsec:col_calib}, while ${C_{\rm corr}}$ refers to the correction (constant offset) applied to the observed colour as derived in Sec.\,\ref{subsec:col_calib}. When calculating the reddening in UV, only the median of the white dwarf distribution was considered to avoid the outliers that deviate from the majority of sources in the colour-colour plot (either from large or low surface gravities, binarity, or data issues), and we followed the same $\chi^2$ minimisation method as in Sec.\,\ref{subsec:col_calib}. A bootstrapping technique was used to estimate the reddening and errors in each bin to account for uncertainties arising from binning or photometric measurements.

\begin{table}
\caption{\textit{GALEX} sample selection of DA white dwarfs for reddening study.}
\centering
\begin{tabular}{cc}
\hline
Criteria & Number of objects \\\hline
\textit{Gaia}-\textit{GALEX} cross-match & 11\,775\\
$P_{\rm WD}$ $\geq$ 0.75 & 11\,455\\
Distance $\leq$300\,pc & 4115 \\
$G, G_{\rm BP}, G_{\rm RP} \leq19$ mag & 2594 \\
$13\,000 \leq\Teff \leq 40\,000$ K & 1458 \\
$FUV, NUV \leq20.5$\,mag & 1261\\
$\av\leq0.1$\,mag & \textbf{1158}\\
\hline
\end{tabular}
\label{tab:galex_sample}
\end{table}

\subsection{Results}\label{sec:galex_res}
\subsubsection*{\center UV Reddening variation with $A_{V}$ and distance}
We used bin widths of 20\,pc and 0.01\,mag in distance and $A_{V}$, respectively, to calculate the reddening in each bin for \textit{NUV$-$G} and \textit{FUV$-$G} colours using the $A_{V}^{\rm map}$ values (Sec.\,\ref{sec:avmap}).

\begin{figure}
\centering
\includegraphics[width=\columnwidth]{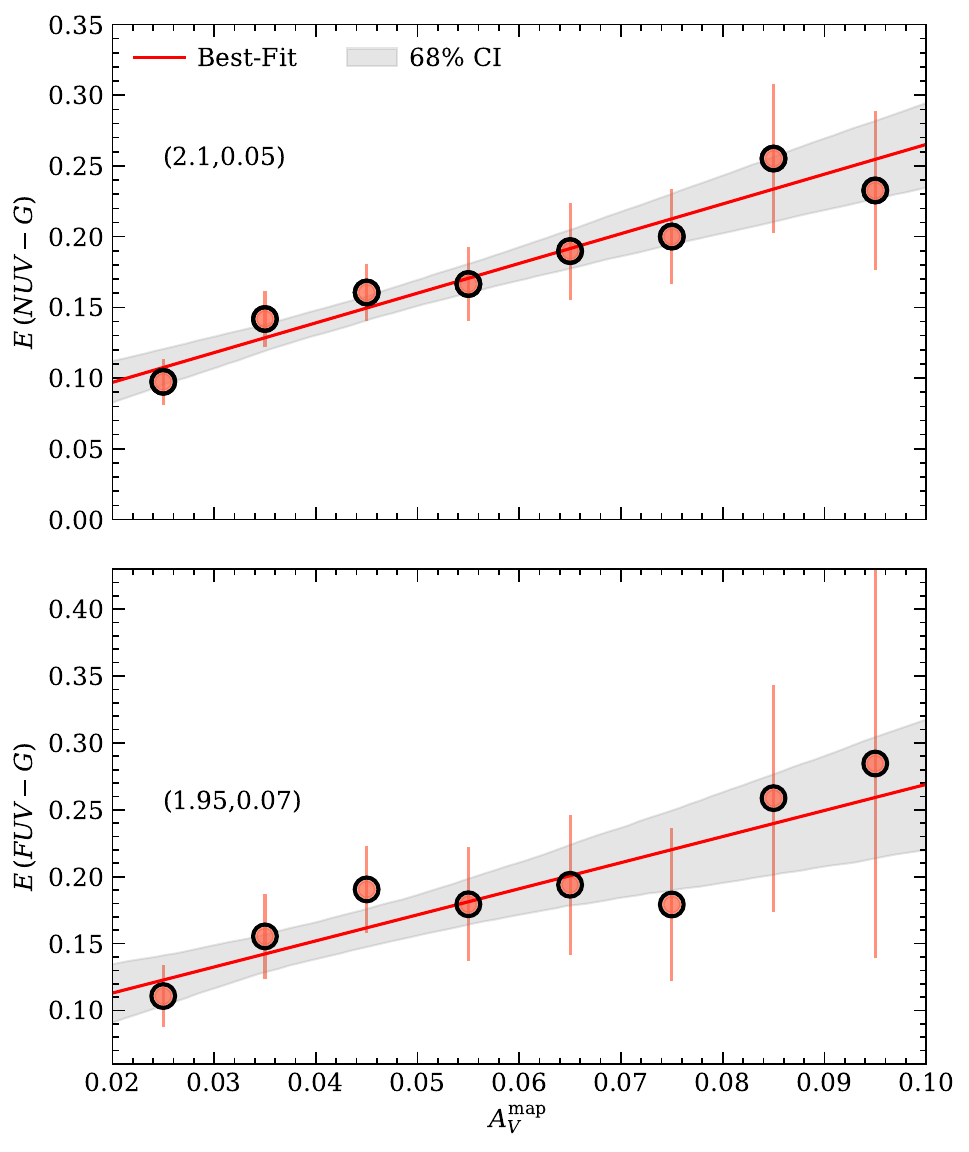}
\caption{UV reddening as a function of optical extinction ($A_{V}^{\rm map}$) in \textit{NUV$-$G} and \textit{FUV$-$G} as shown in top and bottom panels, respectively. The reddening values determined from the comparison of the white dwarf models and observed colours in several bins are marked in light red circles. The red solid line corresponds to the best fit obtained from Bayesian analysis. The filled grey region shows the 68\,per cent confidence interval extracted from 1000 iterations. The slope and intercept of the best fit are labeled in the brackets in the panel.}
\label{fig:Av_bayes_fit}
\end{figure}

\begin{figure}
\centering
\includegraphics[width=\columnwidth]{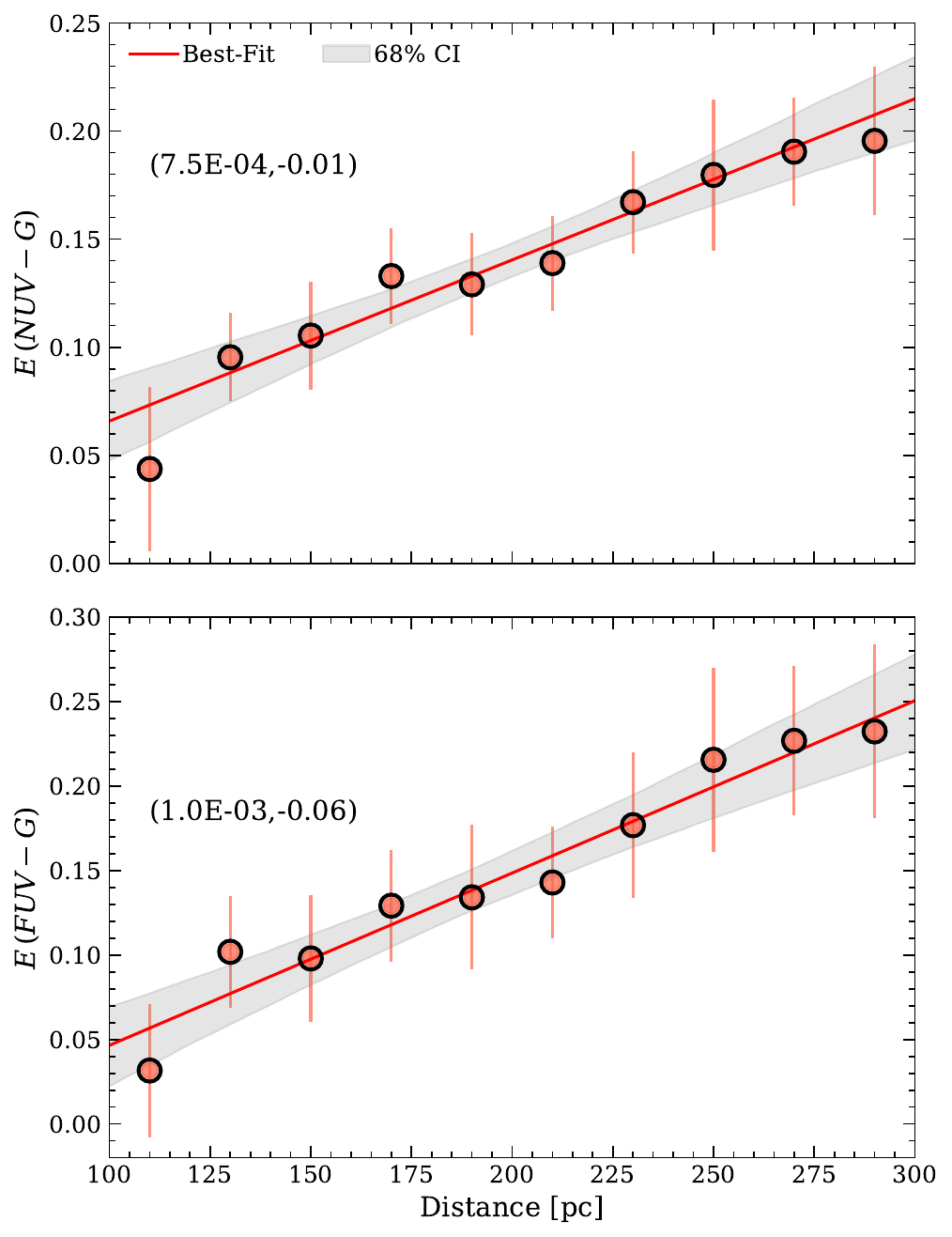}
\caption{Same as in Fig.\,\ref{fig:Av_bayes_fit} but showing the variation with distance extending to 300\,pc in bins of 20\,pc. We find a linear variation of \textit{NUV$-$G} and \textit{FUV$-$G} reddening values with distance. }
\label{fig:av_dist_bayes_fit}
\end{figure}

Figures\,\ref{fig:Av_bayes_fit} and \,\ref{fig:av_dist_bayes_fit} show the UV reddening as a function of $A_{V}^{\rm map}$ and distance with the $x$-axis being the average of the $A_{V}$/distance range in each bin. We observe that the UV reddening varies linearly with distance and $A_{V}^{\rm map}$, where we find large uncertainties at higher extinction ($A_{V}^{\rm map}>0.08$\,mag). This might be due to low number statistics arising from the sudden drop (from 45 to $<20$\,per cent) in the number of sources with \textit{GALEX} magnitudes brighter than 17\,mag in these high extinction bins.

\begin{figure}
\centering
\includegraphics[width=\columnwidth]{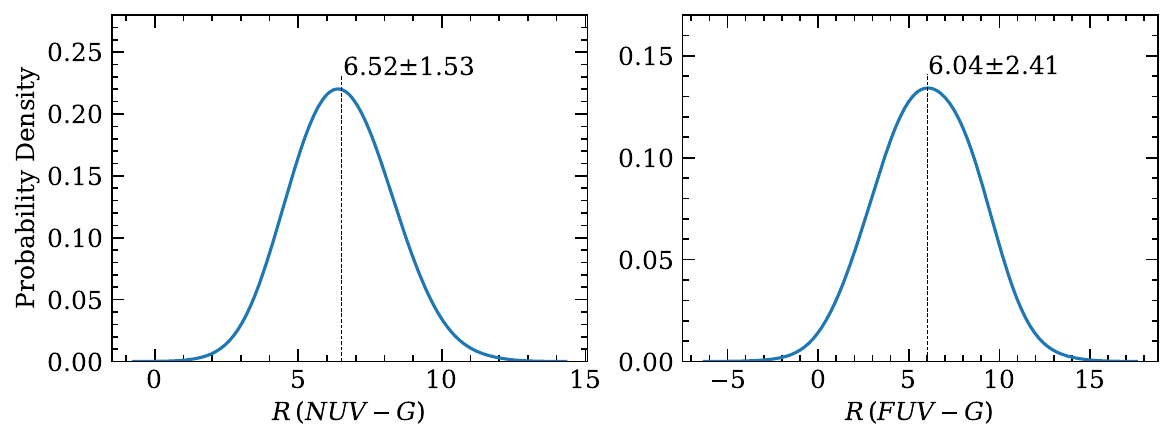}
\caption{Probability distribution function of the best-fit reddening coefficients $R(NUV-G)$ (left panel) and $R(FUV-G)$ (right panel) to the ensemble of stars lying in the \Teff~range 13\,000--40\,000\,K. The reddening values and corresponding 1$\sigma$ errors obtained in Eqns.\,\ref{eqn:redd_calc3}-\ref{eqn:redd_calc4} are labeled in the figure.}
\label{fig:redd_coeff}
\end{figure}

To determine the linear reddening coefficient $R(F/NUV-G)$, we fitted the measured $E(F/NUV - G)$ (Eqns.\,\ref{eqn:redd_calc1} \& \ref{eqn:redd_calc2}) as:
\begin{align}
    E(F/NUV - G)= R(F/NUV-G) \times E(B-V)^{\rm map} \\=R(F/NUV-G) \times A_{V}^{\rm map} / 3.1.
\end{align}
using Bayesian analysis from python \texttt{pymc} package. To get more realistic errors in the linear fit, we performed 1000 iterations to generate a posterior Gaussian distribution of the best-fit parameters, and computed the 68\,per cent confidence interval representing the 1$\sigma$ errors. The best-fit linear relations with $A_{V}^{\rm map}$ are given below:
\begin{align}
E(NUV-G)=2.10\,(\pm0.50)\times A_{{\rm V}}^{\rm map} +0.06\,(\pm0.02)\\
E(FUV-G)=1.95\,(\pm0.78)\times A_{V}^{\rm map} +0.07\,(\pm0.03)
\end{align}
which is equal to 
\begin{align}
    E(NUV-G)=6.52\,(\pm1.53) \times E(B-V)^{\rm map}+0.06\, (\pm0.02)\label{eqn:redd_calc3}\\
    E(FUV-G)= 6.04\,(\pm 2.41) \times E(B-V)^{\rm map}+0.07\,(\pm 0.03)\label{eqn:redd_calc4}
\end{align}
Thus, the reddening coefficients $R(NUV-G)$ and $R(FUV-G)$ are found to be $6.52\,(\pm1.53)$ and $6.04\,(\pm 2.41)$ respectively. Their posterior distributions are shown in Fig.\,\ref{fig:redd_coeff}. The $y$-intercepts obtained from the linear fitting are subjected to the uncertainties in the \textit{GALEX} flux calibration, $A_{V}^{\rm map}$ bias and white dwarf models. Nevertheless, the derived reddening coefficients which correspond to the slope of the correlation are not affected by these biases.

From the observed fits shown in Figs.\,\ref{fig:Av_bayes_fit} \& \ref{fig:av_dist_bayes_fit}, it is evident that $E(FUV-G)$ and $E(NUV-G)$ correlate well with optical extinction and distance. Extrapolating the fits in Eqns.\,\ref{eqn:redd_calc3} \& \ref{eqn:redd_calc4} (Fig.\,\ref{fig:Av_bayes_fit}) to low extinctions ($\av<0.02$\,mag), this suggests an  $A_{V}^{\rm map}$ bias of $-0.03\,(\pm 0.01)$\,mag. This is roughly in agreement within error bars with the $A_{V}^{\rm map}$ map bias found in our \textit{HST} analysis.

The measured reddening values depend explicitly on $A_{V}^{\rm map}$ values that are used for dereddening \textit{Gaia} colours, hence this significantly affects the slope and the derived UV reddening coefficients. To check for this effect, we assumed that the \textit{GALEX} reddening remains constant with $A_{V}^{\rm map}$ (same as $C_{{\rm corr}, F/NUV}$) to estimate a minimum value (lower limit) for $R(F/NUV-G)$. This leads to lower limits for UV reddening coefficients of $\approx5.2$ in both the \textit{GALEX} filters.

\section{Discussion}\label{sec:discuss}
With a primary goal to test 3D dust extinction maps using ultraviolet extinction, we discuss and compare below the extinction results obtained from two independent methods (\textit{HST} and \textit{GALEX}) with respect to the integrated extinctions from 3D EXPLORE dust maps and verify their consistency with literature estimates. 

\begin{table*}
\caption{UV reddening coefficients determined in this work from \textit{HST} and \textit{GALEX} compared to different studies.}
\centering
\begin{tabular}{ccccc}
\hline
Study & $R(NUV-G)$  &  $R(FUV-G)$ & $R(NUV)$ & $R(FUV)$ \\\hline
\noalign{\smallskip}
\textit{GALEX}+$A_{V}^{\rm map}$ ($\leq0.1$\,mag) (this work) & \textbf{6.52 (1.53)} & \textbf{6.04 (2.41)} & &\\ 
\textit{GALEX}+$A_{V}^{\rm GF21}$ (this work) & 6.15 (2.32) & 5.74 (3.7)  & &\\
\textit{HST} COS+$A_{V}^{\rm map}$ (this work) &   & 5.14 (1.21)* \\
\noalign{\smallskip}
\hline
\cite{bianchi2011b} & 5.10&  5.21 & 7.95 & 8.01\\
\cite{yuan2013} & 4.39 & 2.04 & 7.24 (0.08) & 4.29 (0.6)\\
\cite{Sun2018} & 4.04  & 3.53  & 6.89 (0.07) & 6.38 (0.16) \\
\cite{wall2019} &  3.87   &  5.16  & 6.72 (0.04) & 8.05 (0.07) \\
\cite{zhang2023}& 4.44 &  4.12 & 7.29 & 6.97  \\
\hline
\end{tabular}
\footnotesize\flushleft where * denotes the value derived by convolving the COS G130M filter throughput with a white dwarf model spectrum of \Teff$=$20\,000\,K considering an $R(V)=3.3\,(\pm1.2)$ \citep{gordon2023} based on the Si\,II column densities. 
We used $R(G)=2.85$ \citep{gordon2023} to transform UV extinction coefficients from previous studies to reddening coefficients for comparison with this work. $A_{V}^{\rm map}$ refers to current EXPLORE 3D extinction density maps (Sec.\,\ref{sec:avmap}) and $A_{V}^{\rm GF21}$ to the older EXPLORE maps used in \citet{nicola2021}.
\label{tab:redd_coeffs_comp}
\end{table*}

\subsubsection*{\av\ correlations between UV and 3D dust maps}
Based on the \textit{HST} spectroscopic study, the correlation between \av\ from 3D dust maps with those inferred from \ion{Si}{ii} column densities suggest a higher than average gas-to-dust ratio between Galactic longitudes 200 to 300$^{\circ}$ (Fig.\,\ref{fig:rv_galc}). This region falls in the well-known ISM cavity being distinguishable in the 3rd quadrant of the 3D dust maps \citep{lallement2015}. This giant cavity ($\geq1000\,$pc) contains a large fraction of warm and fully ionised gas which is poor in dust and whose origin remains uncertain. In a recent study, \cite{cox2024} reported a high ratio between the density of carbonaceous macromolecules at the origin of the diffuse interstellar band measured by \textit{Gaia} and the dust extinction, also suggesting a high gas-to-dust ratio. The above findings are compatible with our results validating the use of \ion{Si}{ii} for estimating extinction.

A potential explanation for the noted bias between UV and optical extinction correlations could be the combination of the target selection bias effect and the limited resolution of the 3D maps. In dust cloud groups, there are presence of holes or tunnels with lower than average extinction \citep{lallement2015, vergely2022}. These are directions where targets are less attenuated and, hence are easier to detect. In these cases, the 3D maps may overestimate the true extinction because of the smoothing of the opacity that erases the signature of holes while \ion{Si}{ii} measurements represent the true absorption. This effect is prominent at the low Galactic latitude regions ($-35^{\circ}$ to $+45^{\circ}$) towards opaque areas. Hence, the targets located in these regions with extremely low \ion{Si}{ii} column densities may have an overestimated optical extinction. Furthermore, the targets in the lowest \av\ regions suffer the most from overestimation (due to smoothing in the higher \av\ regions located around them). As \av\ increases it becomes less biased in the 3D maps as both higher and lower stellar extinction measurements contribute in the smoothing.

\subsubsection*{\centering Comparison of HST COS and \textit{GALEX} results}
\begin{figure}
\centering
\includegraphics[width=\columnwidth]{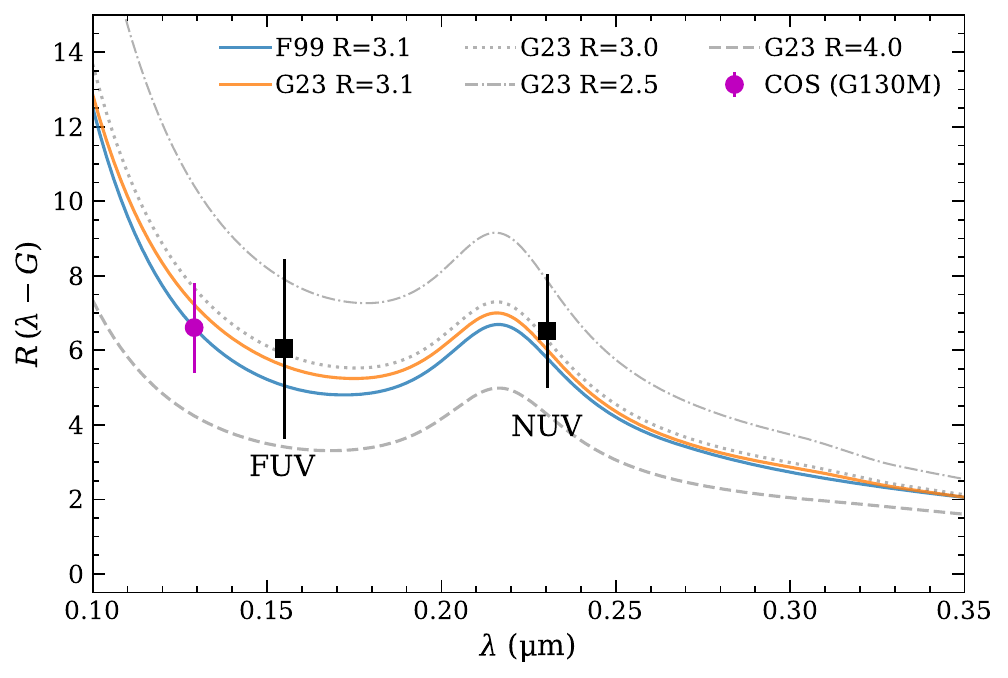}
\caption{Reddening coefficient as a function of wavelength in the UV region. The extinction law curves from \citet{Fitz1999} and \citet{gordon2023} for $R(V)$ values of 2.55, 3.0, 3.1 and 4.0 are shown for comparison with the values $R(\lambda-G)$ derived in this work for \textit{HST} COS and \textit{GALEX} \textit{FUV} and \textit{NUV} filters. Black squares denote the values obtained considering \textit{GALEX} samples with $A_{V}^{\rm map}\leq0.1$\,mag.  For comparison, the reddening coefficient for the COS G130M grating (1291\,\AA) is shown as a magenta dot corresponding to a white dwarf model spectrum \citep{Tremblay2011} of $\Teff= 20\,000$\,K and $R(V) =3.3$.}
\label{fig:uv_ext_laws}
\end{figure}

\textit{HST} COS extinction results are based on \av~measurements from \ion{Si}{ii} lines of DA white dwarfs obtained with G130M spectra that lie in the \textit{FUV} region (1130-1430\,\AA) and within the wavelength range of \textit{GALEX} FUV. Considering a white dwarf model atmosphere of \Teff\ = 20\,000\,K, we convolved it with the COS G130M filter throughput to derive $R(COS-G) = 6.6$ (Fig.\,\ref{fig:uv_ext_laws}), integrating it with an extinction law \citep{gordon2023} with a median $R(V)=3.3$ ($\pm1.2$) obtained from the extinction correlations (Sec.~\ref{sec:hst_res}). This is equivalent to $R(FUV-G)= 5.14$, considering the \textit{GALEX} \textit{FUV} bandpass.
Comparing with the observed \textit{GALEX} \textit{FUV} reddening coefficient, we find $R(FUV-G)=6.04$ ($\pm2.41$) from the linear relation with $A_{V}^{\rm map}$ (Table\,\ref{tab:redd_coeffs_comp}) indicating that the results are in reasonable agreement with COS (within 1$\sigma$ error). 

Extinction correlations of \textit{GALEX} with $A_{V}^{\rm map}$ suggest an average bias on the order of 0.03\,mag, roughly consistent with the \textit{HST} results. Given the large uncertainties associated with the variations of \textit{GALEX} reddening with \av, the noted bias is not well constrained. Compared to this, the \textit{HST} results based on individual white dwarfs are more robust since a bias is still noted considering the effect of dust depletion. Using \textit{HST}, we obtained a bias of 0.01\,mag with $A_{V}^{\rm map}$ being higher than \av\ from \ion{Si}{ii} columns. A similar bias is noted by \cite{vergely2022}, where they found the \av\ values from the EXPLORE 3D maps to be higher than extinctions measured using \ion{P}{ii} and other UV absorption line columns of 31 nearby WDs located within 200\,pc \citep{Lehner2003}. However, in our case, we note that the bias is dependent on the Galactic latitude with the high latitude regions ($>+30^{\circ}$) having a larger value of 0.02\,mag. We investigated if this trend in bias with Galactic latitude is still noted considering a different dust extinction map from \cite{leike2020}. Their maps cover distances up to 400\,pc and are of spatial resolution 2\,pc, higher than the 5\,pc resolution for distances up to 750\,pc used in this work. We derived the \av\ values by integrating the 3D dust extinction maps of \cite{leike2020} along the distance and Galactic position, similarly to the EXPLORE maps. Comparing $A_{V}^{\ion{Si}{ii}}$ with their values (Fig.\,\ref{fig:av_si_leike20}), we find a weaker correlation for lower Galactic latitude regions (from $-20^{\circ}$ to $+20^{\circ}$) while showing larger \av\ disagreements for the higher Galactic latitude regions ($\geq35^{\circ}$) than EXPLORE maps. On the other hand, comparing the targets with depletion measurements (Fig.\,\ref{fig:av_nh_leike20}), we find the \av\ values to be in better agreement with a bias of 0.01\,mag with $A_{V}^{\ion{Si}{ii}}$ being lower than $A_{V}^{\rm map}$, especially for cases with log\,$N({\rm H})<19.5$.

\subsubsection*{Comparison of \textit{GALEX} reddening coefficients with previous studies}
\cite{bianchi2011b} theoretically computed the extinction coefficients using the \cite{Cardelli1989} extinction law for the Milky Way. Based on observations, \cite{yuan2013} derived the UV extinction coefficients by combining \textit{GALEX} and 2MASS photometry of main-sequence stars spanning the \Teff~range 5000--8000\,K. They found a large deviation in the \textit{FUV} extinction measurement when compared with the theoretical values derived from extinction laws \citep{Cardelli1989, Fitz1999}. Later, \cite{Sun2018} calculated the UV colour excess by combining \textit{GALEX} and APASS photometry with a spectroscopic sample of 25\,000 A and F-type stars and found that both \textit{FUV} and \textit{NUV} extinction results are in general agreement with the theoretical prediction for $R(V)=3.35$ \citep{Fitz1999}. In a recent study, \cite{zhang2023} measured these values and studied their dependence on stellar spectral energy distribution and optical extinction, finding good agreement with theoretical predicted values. Comparing the absolute and synthetic magnitudes of white dwarfs, \cite{wall2019} reported a lower value than all the above-reported studies, with $FUV$ results agreeing well with \cite{bianchi2011b}. 

\begin{figure}
\centering
\includegraphics[width=\columnwidth]{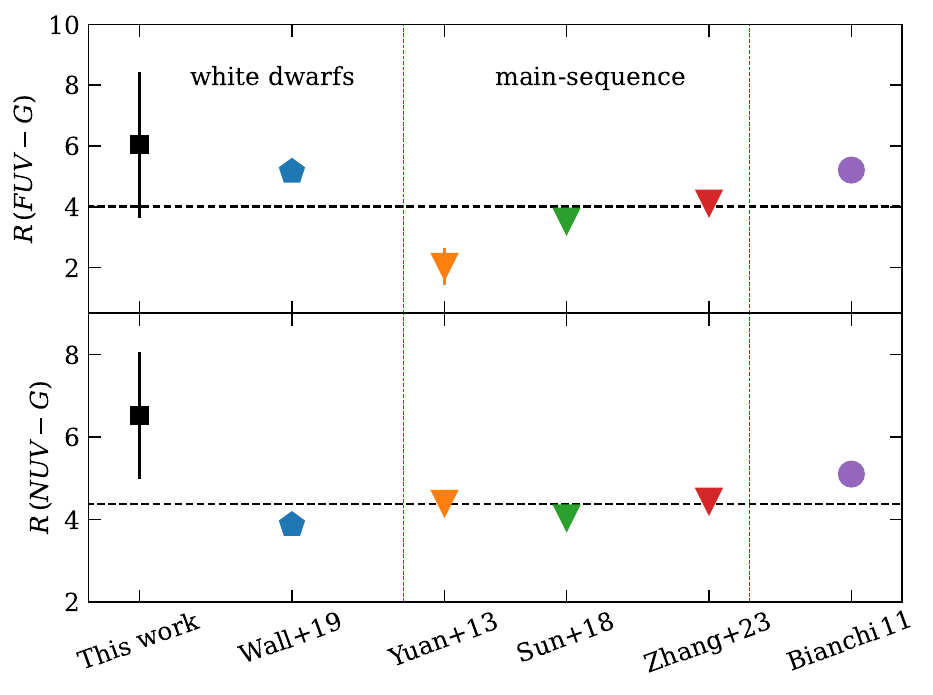}
\caption{Comparison of UV reddening coefficients derived in this work with reported values in the literature ($x$-axis). The average of the published studies excluding this work is shown as a black horizontal dashed line. Refer to Fig.\,\ref{fig:uv_ext_laws} for details of the symbols used for this work. The magenta dotted line separates the results based on white dwarfs and main-sequence stars while the green dotted line separates these results from \citet{bianchi2011b} based on a theoretical extinction law for the Milky Way \citep{Cardelli1989}. The comparison is with the studies from \citet{wall2019} (Wall+19), \citet{yuan2013} (Yuan+13), \citet{Sun2018} (Sun+18),  \citet{zhang2023} (Zhang+23), and \citet{bianchi2011b} (Bianchi\,11)}
\label{fig:uv_redd_comp}
\end{figure}

We transformed the extinction coefficients of the aforementioned studies in terms of $R(FUV-G)$ and $R(NUV-G)$ (Table\,\ref{tab:redd_coeffs_comp}) to compare them with the values derived in this work. During the conversion, the extinction coefficient in \textit{Gaia} $G$ band ($R_{G}=2.85$) was derived considering the empirical extinction law of \cite{gordon2023} at the effective wavelength (5822\,\AA) for reference. The comparisons are shown in Fig.\,\ref{fig:uv_redd_comp}. Given the uncertainties in our measurements, we find that our values are 1.5 times higher than previous studies having an average $R(FUV-G)$ and $R(NUV-G)$ values of 4.01 and 4.37 in \textit{FUV} and \textit{NUV} bands respectively. Specifically, we find $R(FUV-G)$ to be in better agreement with \cite{wall2019} and \cite{bianchi2011b} than the results based on main-sequence stars. 

\subsubsection*{\centering Comparison with \cite{nicola2021}}
The widely used \citet{nicola2021} catalogue of white dwarfs provides \textit{Gaia}-derived 
\Teff, $\log g$ and masses for which the observations are dereddened using different extinction (
$A_{V}^{\rm GF21}$) values. $A_{V}^{\rm GF21}$ values are based on 3D maps of a 10\,pc spatial resolution compared to a higher resolution 5\,pc map used in this work for targets within 750\,pc. We examined the UV reddening variation with $A_{V}^{\rm GF21}$ values instead of the updated values from EXPLORE. For \textit{GALEX} data, a linear fit of $E(NUV-G)$ and $E(FUV-G)$ with $A_{V}^{\rm GF21}$ was performed (with an increased bin size of 0.015). We found a similar correlation as that with $A_{V}^{\rm map}$: $E(NUV-G)=1.98\times A_{V}^{\rm GF21}+0.07$ and $E(FUV-G)=1.85\times A_{V}^{\rm GF21}+0.09$.
Considering these fits, the derived reddening coefficient in $FUV$ (Table\,\ref{tab:redd_coeffs_comp}) shows a larger deviation from the extinction laws compared to the results with $A_{V}^{\rm map}$ (Sec.\,\ref{sec:galex_res}), where we find a better agreement. The differences are not statistically significant though given the large uncertainties in the measurements. Since the updated EXPLORE map is based on the \textit{Gaia}\,EDR3 catalogue that provides better accuracy in parallaxes and reddening coefficients than the earlier maps based on DR2, it is recommended to use the version that is publicly available via G-Tomo application on the EXPLORE website for extinction corrections. 

\section{Conclusion}\label{sec:conc}
In this work, we studied UV extinction using spectroscopic and photometric observations of DA white dwarfs. To achieve our primary goal of validating the latest 3D dust extinction maps from EXPLORE \citep{vergely2022} based on optical observations and investigating bias in optical extinction, we performed several experiments in the UV regime (1140--2830\,\AA). For the first time, we utilised \textit{HST} COS spectra of 180 white dwarfs located within 200\,pc to infer gas column densities towards these sight lines. Selecting 76 sight lines with well-measured ISM parameters from \Lines{Si}{ii}{1190, 1193, 1260} lines, we investigated the variation of column densities and resulting optical extinction as a function of integrated extinction from 3D dust maps. We found the extinctions from \ion{Si}{ii} columns being 0.01--0.02\,mag lower than those based on 3D maps and varying with the Galactic latitude. Strikingly, the observed central velocities are noted to be in good agreement with the predicted velocities from the LISM model \citep{seth2008}, thus providing additional validation of the model and extinction results. We identify several candidates with either larger velocity differences than the model predictions ($>{12}$\,km\,s$^{-1}$) or with a higher velocity dispersion ($b>5$\,km\,s$^{-1}$) suggesting the presence of multiple clouds along the line of sight. These targets require follow-up with high-resolution UV spectroscopic observations to resolve the individual clouds. We derived the synthetic depletion ($F_{*}$) and log\,$N$(H) for 16 sight lines using the column density measurements of several other UV ions (\ion{C}{ii}, \ion{N}{i}, \ion{S}{ii}, \ion{P}{ii}, \ion{Fe}{ii}) present in the COS spectra, showcasing the capabilities achievable with moderate-resolution spectra. Our results probe the depletion in different lines of sight within the Local Bubble and are consistent with \cite{Jenkins2009} measurements. Studying the \av\ correlations with Galactic longitude, we find that \ion{Si}{ii} serves as an appropriate metric for estimating extinction. Our study with the COS sample is a substantial improvement over previous works, in terms of the number of sight lines for constraining the bias in optical extinction.

In addition, we verified 3D dust maps by matching white dwarf models with photometric colours obtained from a \textit{GALEX-Gaia} sample of 1158 DA white dwarfs. We first calibrated the \textit{GALEX} magnitudes and colours utilising the COS sample and white dwarf model spectra. Using colour-colour diagrams, UV reddening variation was studied as a function of optical extinction and distance. We noted a linearly increasing trend in the UV reddening as a function of optical extinction with $A_{V} \leq$\,0.1\,mag and distances up to 300\,pc for the entire \Teff~range (13\,000--40\,000\,K) in \textit{NUV} and \textit{FUV} bands. These correlations demonstrate that \textit{GALEX} photometric observations are reliable for UV extinction measurements. We derived reddening coefficients in \textit{GALEX} bands of $R(FUV-G) = 6.04\pm2.41$ and $R(NUV-G) = 6.52\pm1.53$, respectively. We found that these reddening coefficients are higher than those found in previous studies not based on EXPLORE 3D maps while being consistent with the extinction laws \citep{Fitz1999, gordon2023}. \textit{HST} and \textit{GALEX} data independently demonstrate the reliability of 3D maps from EXPLORE for extinction and reddening corrections of white dwarfs at UV and optical wavelengths. Our study is limited to optical extinction values less than 0.1\,mag due to a reduction in the number of white dwarfs with reliable \textit{GALEX} photometry beyond 300\,pc. Future UV spectroscopic and photometric surveys of white dwarfs with better sensitivity than \textit{GALEX} such as Ultraviolet Explorer (\textit{UVEX}; \citealt{uvex2021}) will have the capability to validate 3D maps out to kilo-parsec distances.

\section*{Acknowledgements}
We thank the anonymous referee for valuable suggestions that has improved the manuscript considerably. This research is based on observations made with the NASA/ESA Hubble Space Telescope obtained from the Space Telescope Science Institute, which is operated by the Association of Universities for Research in Astronomy, Inc., under NASA contract NAS 5–26555. These observations are associated with programs 12169, 12474, 13652, 14077, 15073, 16011, 16642, and 17420. SS and PET received funding from the European Research Council under the European Union’s Horizon 2020 research and innovation programme number 101002408 (MOS100PC). This project has received funding from the European Research Council (ERC) under the European Union’s Horizon 2020 Framework Programme (grant agreement no. 101020057).

\section{Data availability}
The COS spectroscopy data underlying this paper are available in the raw form via the \textit{HST} MAST archive under the programs mentioned in the acknowledgements. The \textit{GALEX-Gaia} crossmatched data utilised in this work will be made available via the VizieR Service for Astronomical Catalogues. All additional data underlying this paper are publicly available from the relevant survey archives or will be shared on reasonable request to the corresponding author.

\bibliographystyle{mnras}
\bibliography{ref}

\appendix
\section{ISM parameters of \textit{HST} COS sample}
Table\,\ref{tab:ism_param}-\ref{tab:ism_param1} provides the ISM parameters of 76 sight lines used for comparison with \av\ from 3D maps. Table\,\ref{tab:ism_param_dep} provides the synthetic depletion and log $N$\,(H) derived from the column densities of \ion{C}{ii}, \ion{N}{i}, \ion{Si}{ii}, \ion{P}{ii}, \ion{S}{ii}, \ion{Fe}{ii} following \cite{Jenkins2009} method.

\begin{table*}
\caption{ISM parameters of 76 sight lines derived from \ion{Si}{ii} interstellar lines and used for \av\ comparison. The table is available online through Vizier.}
\centering
\addtolength{\tabcolsep}{-4pt}
\begin{tabular}{cccccccccccc}
\hline
Name	& longitude &	latitude	&	distance	&	$A_{V}^{\rm map}$ & $V_{\rm LISM}$ & $\chi^{2}_{r}$	&	$b$  &	$V_{\rm c}^{\ion{Si}{ii}}$		 	&	log\,$N$(\ion{Si}{ii})	 &	 	 log\,$N$(H)$_{\rm syn}$	&	$A_{V}^{\ion{Si}{ii}}$ \\
& [deg] &	[deg] &	[pc]	& [mag]	&	 [km\,s$^{-1}$] & & [km\,s$^{-1}$]&	[km\,s$^{-1}$]	&	& & [mag]	 \\\hline
APASSJ152827.83$-$251503.0	&	342.24	&	25.39	&	51.79	&	0.05	&	$-$25.58	(0.5)	&	0.66	&	3.65	(1.64)	&	$-$29.30	(1.79)	&	14.86	(0.57)	&	19.47	&	0.015	(0.009)	\\
HE0131+0149	&	144.18	&	$-$59.01	&	47.81	&	0.01	&	13.50	(0.42)	&	0.55	&	4.22	(1.82)	&	14.13	(1.44)	&	14.74	(0.81)	&	19.35	&	0.012	(0.009)	\\
HE0308$-$2305	&	213.14	&	$-$58.18	&	55.74	&	0.01	&	16.15	(0.78)	&	2.65	&	1.65	(1.15)	&	22.26	(0.9)	&	14.95	(0.09)	&	19.56	&	0.019	(0.002)	\\
HE1518$-$0020	&	1.67	&	44.57	&	70.77	&	0.05	&	$-$21.36	(0.56)	&	1.02	&	4.65	(1.34)	&	$-$29.85	(1.31)	&	15.21	(0.33)	&	19.82	&	0.034	(0.011)	\\
HS1334+0701	&	332.74	&	67.00	&	105.82	&	0.03	&	$-$14.69	(0.29)	&	0.80	&	1.50	(2.35)	&	$-$9.07	(1.32)	&	14.85	(0.27)	&	19.46	&	0.015	(0.004)	\\
HS2056+0721	&	55.82	&	$-$23.91	&	79.96	&	0.04	&	$-$13.32	(0.35)	&	4.49	&	2.88	(0.66)	&	$-$21.13	(0.8)	&	14.94	(0.2)	&	19.55	&	0.018	(0.004)	\\
HS2210+2323	&	82.19	&	$-$26.43	&	89.77	&	0.04	&	$-$5.27	(0.31)	&	3.48	&	2.66	(0.7)	&	$-$2.23	(0.88)	&	15.16	(0.12)	&	19.77	&	0.031	(0.004)	\\
PG0821+633	&	153.16	&	34.57	&	97.23	&	0.02	&	12.80	(0.42)	&	1.18	&	2.42	(1.94)	&	6.08	(1.45)	&	15.17	(0.17)	&	19.78	&	0.031	(0.005)	\\
PG0846+558	&	161.93	&	38.66	&	202.06	&	0.04	&	13.02	(0.42)	&	6.71	&	4.77	(1.17)	&	3.02	(0.85)	&	14.89	(0.52)	&	19.50	&	0.016	(0.009)	\\
PG0915+526	&	165.12	&	43.40	&	72.80	&	0.02	&	11.91	(0.43)	&	0.68	&	1.50	(2.9)	&	10.25	(3.92)	&	14.15	(0.76)	&	18.76	&	0.003	(0.002)	\\
PG1220+234	&	243.46	&	82.47	&	92.89	&	0.02	&	$-$0.59	(0.24)	&	1.14	&	2.12	(0.56)	&	$-$2.69	(1.12)	&	12.76	(0.2)	&	17.37	&	0.0001	(0.0001)	\\
PG1508+549	&	90.09	&	52.71	&	85.76	&	0.02	&	$-$6.53	(0.3)	&	1.34	&	3.61	(1.12)	&	$-$15.35	(1.29)	&	15.13	(0.23)	&	19.74	&	0.029	(0.007)	\\
PG1620+260	&	44.07	&	43.11	&	143.18	&	0.06	&	$-$23.85	(0.46)	&	6.79	&	2.56	(0.48)	&	$-$25.21	(0.76)	&	15.14	(0.11)	&	19.75	&	0.029	(0.003)	\\
SDSSJ081305.55+140317.4	&	209.04	&	24.32	&	94.15	&	0.02	&	25.05	(0.52)	&	1.10	&	2.33	(0.92)	&	15.32	(1.35)	&	14.90	(0.18)	&	19.51	&	0.017	(0.003)	\\
WD0018$-$339	&	345.79	&	$-$80.75	&	65.56	&	0.01	&	1.94	(0.46)	&	2.27	&	1.50	(0.54)	&	0.97	(0.85)	&	14.99	(0.07)	&	19.60	&	0.021	(0.001)	\\
WD0136+768	&	125.78	&	14.55	&	74.75	&	0.03	&	9.48	(0.44)	&	0.94	&	1.56	(2.6)	&	4.22	(1.41)	&	15.21	(0.11)	&	19.82	&	0.034	(0.004)	\\
WD0220+222	&	149.36	&	$-$35.64	&	78.71	&	0.03	&	16.23	(0.48)	&	0.64	&	3.84	(1.67)	&	5.61	(1.73)	&	15.19	(0.33)	&	19.80	&	0.033	(0.011)	\\
WD0232+525	&	138.57	&	$-$6.92	&	28.84	&	0.02	&	16.00	(0.4)	&	0.83	&	1.50	(1.7)	&	12.74	(2.21)	&	14.16	(0.4)	&	18.77	&	0.003	(0.001)	\\
WD0308+188	&	163.16	&	$-$32.68	&	48.11	&	0.02	&	17.67	(0.45)	&	1.70	&	1.91	(0.73)	&	17.48	(0.94)	&	14.75	(0.16)	&	19.36	&	0.012	(0.002)	\\
WD0316+345	&	154.52	&	$-$18.92	&	48.95	&	0.02	&	20.35	(0.35)	&	0.85	&	1.71	(0.79)	&	14.37	(0.95)	&	14.42	(0.37)	&	19.03	&	0.006	(0.002)	\\
WD0416+334	&	165.15	&	$-$11.67	&	72.71	&	0.03	&	22.25	(0.33)	&	0.96	&	2.69	(1.29)	&	18.09	(1.35)	&	14.69	(0.45)	&	19.30	&	0.01	(0.005)	\\
WD0701$-$587	&	269.15	&	$-$21.61	&	29.39	&	0.01	&	14.81	(0.56)	&	0.48	&	3.21	(2.25)	&	15.33	(2.6)	&	15.19	(0.35)	&	19.80	&	0.033	(0.012)	\\
WD0808+435	&	176.83	&	32.32	&	68.85	&	0.02	&	16.40	(0.39)	&	1.17	&	1.50	(2.1)	&	10.67	(1.26)	&	14.88	(0.22)	&	19.49	&	0.016	(0.004)	\\
WD0947+325	&	194.11	&	50.73	&	71.41	&	0.02	&	10.34	(0.43)	&	1.08	&	2.67	(0.68)	&	7.14	(0.93)	&	13.14	(0.33)	&	17.75	&	0.0003	(0.0001)	\\
WD1049$-$158	&	265.69	&	37.98	&	38.69	&	0.01	&	8.03	(0.64)	&	1.17	&	1.52	(0.86)	&	2.07	(0.97)	&	14.67	(0.18)	&	19.28	&	0.01	(0.002)	\\
WD1058$-$129	&	265.92	&	41.57	&	46.70	&	0.01	&	7.24	(0.64)	&	2.25	&	1.50	(1.2)	&	4.04	(0.73)	&	14.77	(0.17)	&	19.38	&	0.012	(0.002)	\\
WD1102+748	&	131.87	&	40.63	&	51.88	&	0.01	&	6.53	(0.45)	&	1.72	&	2.38	(0.71)	&	$-$0.49	(0.84)	&	14.90	(0.16)	&	19.51	&	0.017	(0.003)	\\
WD1230$-$308	&	298.30	&	31.56	&	132.84	&	0.06	&	$-$6.82	(0.61)	&	6.99	&	2.94	(0.93)	&	$-$2.83	(1.04)	&	15.58	(0.08)	&	20.19	&	0.081	(0.006)	\\
WD1233$-$164	&	297.68	&	46.02	&	66.76	&	0.02	&	$-$0.76	(0.31)	&	2.77	&	2.34	(0.68)	&	$-$5.32	(1.05)	&	15.07	(0.12)	&	19.68	&	0.025	(0.003)	\\
WD1258+593	&	120.71	&	58.02	&	64.53	&	0.01	&	0.31	(0.46)	&	0.52	&	1.50	(2.4)	&	$-$6.15	(1.88)	&	14.85	(0.28)	&	19.46	&	0.015	(0.004)	\\
WD1314$-$153	&	311.85	&	46.83	&	55.61	&	0.02	&	$-$10.73	(0.37)	&	0.98	&	1.50	(2.41)	&	$-$9.15	(1.15)	&	15.01	(0.2)	&	19.62	&	0.022	(0.004)	\\
WD1334$-$160	&	318.52	&	45.16	&	59.77	&	0.03	&	$-$18.72	(0.35)	&	1.72	&	2.71	(1.07)	&	$-$10.72	(1.25)	&	15.23	(0.15)	&	19.84	&	0.036	(0.005)	\\
WD1518+636	&	99.94	&	46.46	&	55.48	&	0.02	&	$-$3.13	(0.46)	&	0.76	&	1.50	(1.09)	&	$-$10.16	(1.84)	&	14.64	(0.23)	&	19.25	&	0.009	(0.002)	\\
WD1525+257	&	39.45	&	55.12	&	84.69	&	0.03	&	$-$16.92	(0.37)	&	2.06	&	2.21	(0.95)	&	$-$19.02	(1.17)	&	15.16	(0.13)	&	19.77	&	0.031	(0.004)	\\
WD1527+090	&	14.58	&	48.44	&	54.47	&	0.03	&	$-$20.68	(0.65)	&	2.64	&	1.82	(0.92)	&	$-$25.42	(0.92)	&	15.04	(0.12)	&	19.65	&	0.023	(0.003)	\\
WD1531$-$022	&	2.42	&	40.92	&	41.67	&	0.05	&	$-$26.57	(0.52)	&	2.37	&	1.59	(0.99)	&	$-$24.30	(0.83)	&	15.12	(0.09)	&	19.73	&	0.028	(0.003)	\\
WD1547+057	&	14.19	&	42.52	&	88.38	&	0.06	&	$-$22.64	(0.67)	&	4.00	&	2.66	(0.9)	&	$-$25.00	(0.93)	&	15.29	(0.1)	&	19.90	&	0.041	(0.004)	\\
WD1555$-$089	&	0.84	&	32.05	&	52.17	&	0.07	&	$-$28.96	(0.38)	&	0.62	&	1.50	(2.74)	&	$-$27.86	(1.74)	&	15.14	(0.17)	&	19.75	&	0.029	(0.005)	\\
WD1609+044	&	16.57	&	37.14	&	116.69	&	0.08	&	$-$23.49	(0.34)	&	11.86	&	3.43	(0.78)	&	$-$24.66	(1)	&	15.30	(0.13)	&	19.91	&	0.042	(0.006)	\\
WD1619+123	&	26.92	&	38.62	&	55.82	&	0.04	&	$-$22.48	(0.44)	&	0.91	&	1.50	(0.6)	&	$-$28.07	(1.03)	&	14.64	(0.13)	&	19.25	&	0.009	(0.001)	\\
WD1713+332	&	56.44	&	33.52	&	88.87	&	0.03	&	$-$20.41	(0.49)	&	3.57	&	1.50	(2.09)	&	$-$20.24	(0.93)	&	15.26	(0.09)	&	19.87	&	0.039	(0.003)	\\
WD1911+536	&	84.50	&	18.57	&	22.17	&	0.02	&	$-$6.48	(0.45)	&	2.80	&	2.70	(0.8)	&	$-$12.91	(0.99)	&	15.13	(0.14)	&	19.74	&	0.029	(0.004)	\\
WD1919+145	&	49.40	&	0.13	&	19.88	&	0.02	&	$-$24.18	(0.63)	&	1.77	&	3.31	(1.06)	&	$-$27.77	(1.07)	&	15.09	(0.22)	&	19.70	&	0.026	(0.006)	\\
WD2043$-$635	&	332.35	&	$-$36.92	&	82.48	&	0.02	&	$-$17.41	(0.35)	&	2.95	&	2.15	(0.52)	&	$-$16.41	(0.93)	&	15.04	(0.09)	&	19.65	&	0.023	(0.002)	\\
WD2047+372	&	79.06	&	$-$3.99	&	17.59	&	0.02	&	$-$9.17	(0.45)	&	0.77	&	2.00	(1.37)	&	$-$14.05	(1.53)	&	15.03	(0.19)	&	19.64	&	0.023	(0.004)	\\
WD2220+133	&	76.84	&	$-$35.64	&	76.62	&	0.03	&	$-$5.02	(0.33)	&	1.91	&	1.85	(0.72)	&	$-$6.98	(0.8)	&	14.82	(0.15)	&	19.43	&	0.014	(0.002)	\\
WD2341+322	&	106.79	&	$-$28.20	&	18.60	&	0.02	&	6.11	(0.45)	&	0.53	&	2.43	(1.85)	&	3.00	(1.99)	&	14.82	(0.42)	&	19.43	&	0.014	(0.006)	\\
WDJ003310.51+474212.39	&	119.75	&	$-$15.06	&	55.03	&	0.03	&	10.14	(0.44)	&	1.20	&	1.98	(1.7)	&	1.68	(1.56)	&	14.82	(0.28)	&	19.43	&	0.014	(0.004)	\\
WDJ004331.10+470134.30	&	121.53	&	$-$15.82	&	53.53	&	0.02	&	10.82	(0.43)	&	1.48	&	4.12	(0.81)	&	7.92	(0.97)	&	12.99	(0.16)	&	17.60	&	0.0002	(0.0001)	\\
WDJ012813.78$-$530011.30	&	290.60	&	$-$63.23	&	48.49	&	0.01	&	11.17	(0.63)	&	2.19	&	3.10	(0.96)	&	7.50	(1.07)	&	15.24	(0.16)	&	19.85	&	0.037	(0.006)	\\
WDJ012942.65+422818.11	&	130.42	&	$-$19.84	&	83.27	&	0.03	&	13.94	(0.41)	&	7.43	&	3.75	(0.66)	&	4.15	(0.84)	&	15.16	(0.18)	&	19.77	&	0.031	(0.006)	\\
WDJ030146.30+493659.64	&	143.56	&	$-$7.96	&	57.46	&	0.02	&	17.50	(0.38)	&	0.90	&	1.50	(1.52)	&	10.89	(1.23)	&	14.85	(0.19)	&	19.46	&	0.015	(0.003)	\\
WDJ030236.65$-$230151.23	&	212.41	&	$-$60.10	&	79.12	&	0.01	&	18.23	(0.63)	&	3.22	&	2.98	(0.64)	&	20.44	(0.97)	&	15.12	(0.13)	&	19.73	&	0.028	(0.004)	\\
WDJ051613.96$-$701934.93	&	281.28	&	$-$33.39	&	73.91	&	0.04	&	7.90	(0.56)	&	1.90	&	2.97	(0.91)	&	5.16	(1.02)	&	15.31	(0.14)	&	19.92	&	0.043	(0.006)	\\
WDJ055046.46+261220.67	&	182.93	&	$-$0.43	&	94.08	&	0.03	&	23.22	(0.31)	&	2.53	&	2.25	(0.93)	&	14.63	(1.12)	&	15.09	(0.14)	&	19.70	&	0.026	(0.004)	\\
WDJ061000.36+281428.37	&	183.28	&	4.29	&	57.06	&	0.02	&	22.72	(0.32)	&	0.92	&	2.74	(0.87)	&	19.57	(1.38)	&	12.94	(0.33)	&	17.55	&	0.0002	(0.0001)	\\
WDJ072805.02$-$130256.34	&	228.71	&	2.01	&	73.28	&	0.02	&	17.12	(0.39)	&	1.41	&	3.56	(1.03)	&	16.89	(1.15)	&	12.73	(0.15)	&	17.34	&	0.0001	(0.0001)	\\
WDJ081004.00+032926.91	&	218.94	&	19.06	&	64.78	&	0.01	&	18.60	(0.54)	&	1.08	&	2.75	(0.74)	&	15.29	(1.39)	&	13.08	(0.3)	&	17.69	&	0.0003	(0.0001)	\\
WDJ082532.35$-$072823.21	&	230.98	&	17.03	&	35.48	&	0.01	&	14.35	(0.41)	&	0.76	&	1.75	(0.9)	&	14.22	(1.29)	&	14.37	(0.46)	&	18.98	&	0.005	(0.002)	\\
WDJ112401.30$-$505938.44	&	289.19	&	9.53	&	54.95	&	0.02	&	$-$8.65	(0.31)	&	1.47	&	3.41	(1.06)	&	$-$3.24	(1.12)	&	15.00	(0.28)	&	19.61	&	0.021	(0.006)	\\
\hline
\end{tabular}
\label{tab:ism_param}
\end{table*}

\begin{table*}
\caption{Table A1 continued.}
\centering
\addtolength{\tabcolsep}{-4pt}
\begin{tabular}{cccccccccccc}
\hline
Name	& longitude &	latitude	&	distance	&	$A_{V}^{\rm map}$ & $V_{\rm LISM}$ & $\chi^{2}_{r}$	&	$b$  &	$V_{\rm c}^{\ion{Si}{ii}}$		 	&	log\,$N$(\ion{Si}{ii})	 &	 	 log\,$N$(H)$_{\rm syn}$	&	$A_{V}^{\ion{Si}{ii}}$ \\
& [deg] &	[deg] &	[pc]	& [mag]	&	 [km\,s$^{-1}$] & & [km\,s$^{-1}$]&	[km\,s$^{-1}$]	&	& & [mag]	 \\\hline
WDJ150742.03$-$592754.43	&	319.47	&	$-$1.04	&	80.83	&	0.03	&	$-$19.39	(0.36)	&	3.90	&	3.13	(0.56)	&	$-$17.18	(0.93)	&	14.97	(0.16)	&	19.58	&	0.02	(0.003)	\\
WDJ170707.97+222428.34	&	43.44	&	32.38	&	50.31	&	0.03	&	$-$23.24	(0.55)	&	0.76	&	1.57	(1.5)	&	$-$21.90	(1.57)	&	14.49	(0.36)	&	19.10	&	0.007	(0.002)	\\
WDJ174902.45$-$343255.27	&	355.57	&	$-$3.52	&	55.23	&	0.03	&	$-$30.06	(0.56)	&	1.87	&	3.09	(0.85)	&	$-$25.75	(0.93)	&	14.98	(0.25)	&	19.59	&	0.02	(0.005)	\\
WDJ175352.16+330622.62	&	58.67	&	25.74	&	35.70	&	0.03	&	$-$18.42	(0.57)	&	1.38	&	1.95	(0.66)	&	$-$20.22	(0.91)	&	14.90	(0.13)	&	19.51	&	0.017	(0.002)	\\
WDJ180354.33$-$375202.95	&	354.16	&	$-$7.77	&	60.33	&	0.02	&	$-$29.64	(0.37)	&	2.70	&	2.87	(0.87)	&	$-$26.78	(1.02)	&	15.03	(0.19)	&	19.64	&	0.023	(0.004)	\\
WDJ181140.82+282939.19	&	55.25	&	20.64	&	73.06	&	0.05	&	$-$19.73	(0.56)	&	1.85	&	3.86	(0.97)	&	$-$24.54	(0.94)	&	14.86	(0.39)	&	19.47	&	0.015	(0.006)	\\
WDJ191558.47$-$303535.44	&	7.28	&	$-$18.30	&	51.63	&	0.02	&	$-$20.91	(0.35)	&	2.05	&	3.28	(1.02)	&	$-$25.95	(1.1)	&	15.13	(0.21)	&	19.74	&	0.029	(0.006)	\\
WDJ191720.56+445239.38	&	76.27	&	14.51	&	77.00	&	0.03	&	$-$13.08	(0.65)	&	3.99	&	3.79	(0.8)	&	$-$18.33	(1)	&	15.30	(0.17)	&	19.91	&	0.042	(0.007)	\\
WDJ192034.41$-$471529.44	&	350.55	&	$-$24.30	&	69.09	&	0.02	&	$-$13.69	(0.35)	&	2.12	&	2.32	(0.8)	&	$-$24.32	(0.99)	&	15.05	(0.13)	&	19.66	&	0.024	(0.003)	\\
WDJ193124.43+570419.66	&	88.81	&	17.38	&	69.46	&	0.03	&	$-$4.76	(0.45)	&	5.76	&	3.60	(0.78)	&	$-$12.86	(1)	&	15.21	(0.18)	&	19.82	&	0.034	(0.006)	\\
WDJ193955.06+093219.39	&	47.02	&	$-$6.24	&	82.32	&	0.04	&	$-$22.59	(0.63)	&	4.36	&	4.52	(0.95)	&	$-$23.28	(1)	&	15.09	(0.3)	&	19.70	&	0.026	(0.008)	\\
WDJ202359.51$-$422425.85	&	358.36	&	$-$34.45	&	98.80	&	0.03	&	$-$15.74	(0.56)	&	10.52	&	3.35	(0.72)	&	$-$18.30	(0.92)	&	15.17	(0.16)	&	19.78	&	0.031	(0.005)	\\
WDJ204745.04+323922.58	&	75.12	&	$-$6.78	&	59.74	&	0.04	&	$-$9.71	(0.45)	&	1.84	&	4.30	(1.04)	&	$-$14.94	(1.1)	&	14.99	(0.34)	&	19.60	&	0.021	(0.007)	\\
WDJ215229.65+340743.85	&	85.73	&	$-$15.52	&	70.41	&	0.04	&	$-$5.96	(0.3)	&	1.67	&	3.78	(0.91)	&	$-$10.96	(1.15)	&	14.99	(0.27)	&	19.60	&	0.021	(0.006)	\\
WDJ220238.75$-$280942.13	&	21.27	&	$-$52.80	&	65.01	&	0.01	&	$-$12.55	(0.57)	&	2.53	&	1.89	(1.09)	&	$-$10.05	(1.12)	&	15.36	(0.07)	&	19.97	&	0.049	(0.003)	\\
WDJ230840.77$-$214459.60	&	40.27	&	$-$66.05	&	32.77	&	0.01	&	$-$8.34	(0.51)	&	1.52	&	1.50	(2.31)	&	$-$6.39	(1.16)	&	15.14	(0.09)	&	19.75	&	0.029	(0.003)	\\
\hline
\end{tabular}
\label{tab:ism_param1}
\end{table*}

\begin{table*}
\caption{Synthetic $N$(H) and depletion derived following the \citet{Jenkins2009} method. The full table with the column densities of individual elements is available online through Vizier.}
\centering
\begin{tabular}{ccccccc}
\hline
Name	&	longitude [deg]	&	longitude [deg]	&	distance	[pc]&	log\,$N$(H)$_{\rm syn}$ &	$F^{*}$		&	Elements considered	\\\hline
APASSJ152827.83$-$251503.0	&	342.242	&	25.39	&	51.79	&	19.05	(0.98)	&	$-$0.08	(0.84)	&	C, N, Si, S	\\
HE0131+0149	&	144.177	&	$-$59.012	&	47.81	&	18.02	(0.82)	&	$-$0.98	(0.83)	&	C, Si, S	\\
HS2056+0721	&	55.815	&	$-$23.908	&	79.96	&	19.91	(0.14)	&	0.34	(0.2)	&	C, N, Si, S, Fe	\\
PG0846+558	&	161.926	&	38.657	&	202.06	&	19.02	(0.51)	&	$-$0.06	(0.41)	&	C, N, Si, P, S, Fe	\\
PG1220+234	&	243.456	&	82.472	&	92.89	&	17.83	(0.19)	&	0.11	(0.22)	&	C, N, Si, Fe	\\
WD0018$-$339	&	345.789	&	$-$80.747	&	65.56	&	18.91	(0.23)	&	$-$0.59	(0.21)	&	C, N, Si, S	\\
WD0947+325	&	194.109	&	50.732	&	71.41	&	17.43	(0.34)	&	$-$0.6	(0.34)	&	C, N, Si, S, Fe	\\
WD1555$-$089	&	0.841	&	32.051	&	52.17	&	20.29	(0.44)	&	0.4	(0.41)	&	C, N, Si, S, Fe	\\
WD1619+123	&	26.918	&	38.621	&	55.82	&	18.68	(0.81)	&	$-$0.48	(0.72)	&	N, Si, Fe	\\
WD1713+332	&	56.444	&	33.518	&	88.87	&	19.12	(0.69)	&	$-$0.62	(0.61)	&	N, Si, S, Fe	\\
WDJ004331.10+470134.30	&	121.529	&	$-$15.824	&	53.53	&	17.95	(0.21)	&	0.21	(0.22)	&	C, N, Si, S	\\
WDJ072805.02$-$130256.34	&	228.706	&	2.01	&	73.28	&	17.91	(0.23)	&	0.46	(0.24)	&	C, N, Si, S	\\
WDJ193955.06+093219.39	&	47.016	&	$-$6.241	&	82.32	&	18.9	(0.28)	&	$-$0.06	(0.25)	&	C, N, Si, S, Fe	\\
WDJ202359.51$-$422425.85	&	358.36	&	$-$34.454	&	98.8	&	18.67	(0.27)	&	$-$0.35	(0.25)	&	N, Si, P, S, Fe	\\
WDJ204745.04+323922.58	&	75.116	&	$-$6.783	&	59.74	&	18.53	(0.35)	&	$-$0.95	(0.43)	&	C, N, Si	\\
WDJ215229.65+340743.85	&	85.728	&	$-$15.516	&	70.41	&	18.82	(0.49)	&	$-$0.16	(0.43)	&	N, Si, P, S, Fe	\\
\hline
\end{tabular}
\label{tab:ism_param_dep}
\end{table*}

\section{Comparison with \av\ from other 3D maps}
\begin{figure}
    \centering
    \includegraphics[width=\columnwidth]{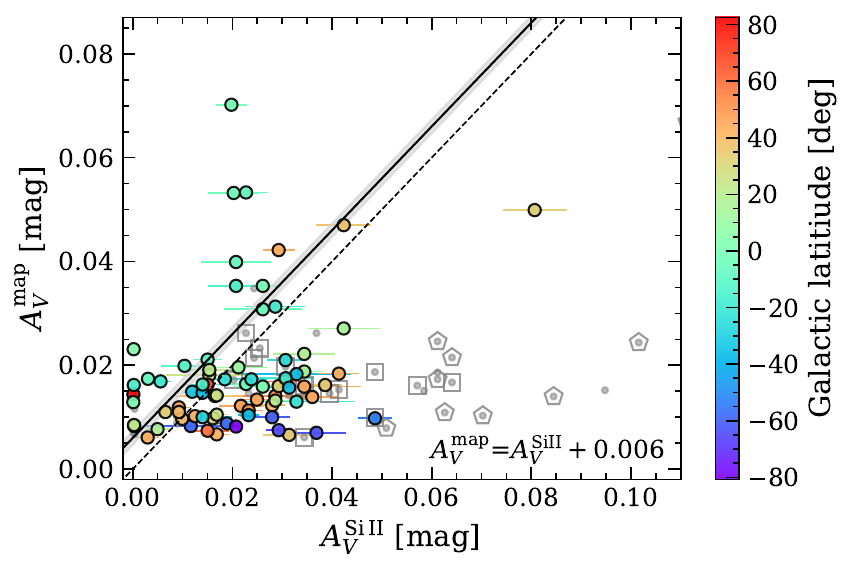}
    \caption{Same as left panel of Fig.\,\ref{fig:ism_Av_comp} but the $A_{V}^{\rm map}$ values are derived using 3D dust extinction maps of \citet{leike2020}. We note that the \av~disagreements are larger for targets located at high Galactic latitudes.}
    \label{fig:av_si_leike20}
\end{figure}

\begin{figure}
    \centering
    \includegraphics[width=\columnwidth]{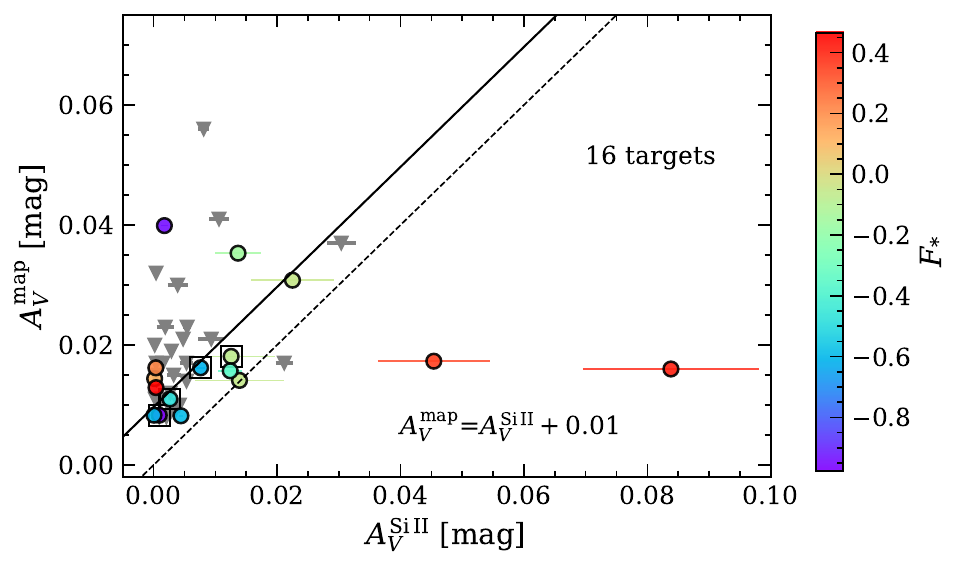}
    \caption{Same as bottom panel of Fig.\,\ref{fig:ism_Av_comp_dep} but with the $A_{V}^{\rm map}$ using 3D dust extinction maps of \citet{leike2020}.}
    \label{fig:av_nh_leike20}
\end{figure}

Figures\,\ref{fig:av_si_leike20} \& \ref{fig:av_nh_leike20} show the comparison of $A_{V}^\ion{Si}{II}$ with the \av\ from \cite{leike2020}.

\end{document}